\documentclass{ecai}

\usepackage{latexsym}
\usepackage{amssymb}
\usepackage{amsmath}
\usepackage{amsthm}
\usepackage{booktabs}
\usepackage{enumitem}
\usepackage{graphicx}
\usepackage[dvipsnames]{xcolor}
\usepackage{color}
\usepackage{subfig}
\usepackage{multirow}

\definecolor{FirstPlace}{rgb}{0.2, 0.6, 0.2}
\definecolor{SecondPlace}{rgb}{0.5, 0.8, 0.2}
\definecolor{ThirdPlace}{rgb}{0.8, 0.8, 0.2}

\newcommand{\first}[1]{\textcolor{FirstPlace}{\textbf{#1}}}
\newcommand{\second}[1]{\textcolor{SecondPlace}{\textbf{#1}}}
\newcommand{\third}[1]{\textcolor{ThirdPlace}{\textbf{#1}}}

\newtheorem{definition}{Definition}

\newcommand{\BibTeX}{B\kern-.05em{\sc i\kern-.025em b}\kern-.08em\TeX}

\begin{document}


\begin{frontmatter}


\paperid{2577} 


\title{Identifying Super Spreaders in Multilayer Networks}


\author[A]{\fnms{Micha{\l}}~\snm{Czuba}\footnote{Equal contribution.}\thanks{Corresponding Author. Email: michal.czuba@pwr.edu.pl}}
\author[A]{\fnms{Mateusz}~\snm{Stolarski}\footnotemark}
\author[B]{\fnms{Adam}~\snm{Pir{\'o}g}}
\author[A]{\fnms{Piotr}~\snm{Bielak}}
\author[A]{\fnms{Piotr}~\snm{Br{\'o}dka}} 

\address[A]{Wroc{\l}aw University of Science and Technology, Wroc{\l}aw, Poland}
\address[B]{4Semantics, Warsaw, Poland}


\begin{abstract}
Identifying super-spreaders can be framed as a subtask of the influence maximisation problem. It seeks to pinpoint agents within a network that, if selected as single diffusion seeds, disseminate information most effectively. Multilayer networks, a specific class of heterogeneous graphs, can capture diverse types of interactions (e.g., physical-virtual or professional-social), and thus offer a more accurate representation of complex relational structures. In this work, we introduce a novel approach to identifying super-spreaders in such networks by leveraging graph neural networks. To this end, we construct a dataset by simulating information diffusion across hundreds of networks --- to the best of our knowledge, the first of its kind tailored specifically to multilayer networks. We further formulate the task as a variation of the ranking prediction problem based on a four-dimensional vector that quantifies each agent's spreading potential: (i) the number of activations; (ii) the duration of the diffusion process; (iii) the peak number of activations; and (iv) the simulation step at which this peak occurs. Our model, \textit{TopSpreadersNetwork}, comprises a relationship-agnostic encoder and a custom aggregation layer. This design enables generalisation to previously unseen data and adapts to varying graph sizes. In an extensive evaluation, we compare our model against classic centrality-based heuristics and competitive deep learning methods. The results, obtained across a broad spectrum of real-world and synthetic multilayer networks, demonstrate that \textit{TopSpreadersNetwork} achieves superior performance in identifying high-impact nodes, while also offering improved interpretability through its structured output.
\end{abstract}

\end{frontmatter}

\section{Introduction}

Spreading phenomena in networks are investigated across various domains, including viral marketing, transportation systems, epidemiology, or social media~\cite{magnani2015spreadingmln}. These studies typically assume the propagation of a certain phenomenon over a graph-like structure simulating a real-world system. A key challenge in this context is to identify the agents that have the greatest impact on the spread. The significance of this task is underscored by its diverse applications, such as detecting vulnerable nodes in power networks, scheduling targeted immunisations, or selecting influencers for advertising campaigns. Nonetheless, addressing such problems requires accurate models of the systems under consideration. To this end, multilayer networks have been proposed~\cite{kivela2014multilayer}. These models provide a more nuanced representation of relationships by enabling, for instance, social ties to be split across multiple layers that correspond to different modes of interaction (e.g., physical, virtual). Considering spreading phenomena within a multilayer setting can better capture the complexity of real-world systems, leading to more precise results.

\subsection{Related works}

We review prior work on techniques to maximise diffusion in complex (particularly multilayer) networks, encompassing both classic and learning-based methods. As a prelude, we first discuss graph neural networks and their capacity to model network dynamics.

\paragraph{Graph neural networks.}

Graph Neural Networks (GNNs)~\cite{gnnSurv} represent a category of deep learning techniques specifically developed for inference on graph-structured data. The standard architecture of GNNs involves an iterative process of exchanging transformed node features and aggregating information from neighbouring nodes (so called message-passing). In a GNN with \textit{k} layers, each node incorporates information from its \textit{k}-hop neighbourhood. Variations in the design of the aggregation function and the combination function give rise to different GNN architectures~\cite{gcn, chiang2019cluster}. The resulting high-level node or graph representations are employed across a range of tasks. GNNs have been successfully applied to problems such as estimating information diffusion~\cite{diff1}, graph-based reasoning~\cite{graphreasoning}, identifying the source of diffusion on graphs~\cite{diff_source1, diff_source2}, or even deep graph generation~\cite{graphgeneration}. By leveraging proper information fusion~\cite{bielak2024representation}, GNNs present a powerful solution to model graph-based data even on multilayer networks.  

\paragraph{Influence maximisation problem.}

The problem of influence maximisation was originally introduced as a combinatorial optimisation task by~\cite{kempe2003maximizing}, laying the foundation for a substantial body of research and practical applications over the following decades. It aims to identify a set of agents which, when used as diffusion seeds, maximise the spread of a given process across the network. The majority of traditional (i.e., non-learning-based) approaches can be classified into three categories: simulation-based, proxy-based, and heuristic-based. 
These methods often rely on centrality-based heuristics, utilising them directly, e.g., selecting top-$k$ nodes with the highest degree centrality, or building upon them, e.g., with discounting algorithms~\cite{chen2009efficient}. Although computationally efficient and interpretable, they operate under local assumptions that limit their adaptability to complex diffusion dynamics. For a comprehensive overview of these techniques, we refer the reader to a survey~\cite{singh2022influence}. 

In recent years, increasing attention has been devoted to addressing influence maximisation problem with learning-based approaches. These methods employ deep learning techniques to overcome the limitations of conventional methods, particularly their poor generalisation capabilities. Within this paradigm, several approaches can be found which model the nodes influence, such as leveraging the latent nodes embeddings~\cite{stolarski2024identifying, kumar2022}, or even incorporate reinforcement learning into influence maximisation frameworks~\cite{Tian2020topicAwareIM, ling2023deepim}. 
For instance, DeepIM~\cite{ling2023deepim} addresses this task through a transductive, generative encoder–decoder architecture augmented with knowledge distillation. While effective, this approach relies on full access to the graph structure during training, which limits its ability.

Nevertheless, most research on influence maximisation continues to focus on the information diffusion within homogeneous network structures, which offer limited expressiveness. More recently, however, attention has begun to shift towards more complex network models, opening new research directions. For instance, Heterogeneous Influence Maximisation incorporates diverse node and edge types by leveraging meta-path-based methods~\cite{Deng2020, mahe2022b}, while Multilayer Influence Maximisation explores information diffusion across multiple platforms via shared nodes~\cite{Venkatakrishna2022CIM}.

As in the case of homogeneous graphs, many approaches in multilayer networks rely on heuristics grounded in a notion of centrality, often adapted from single-layer settings. For instance, neighbourhood-size~\cite{magnani2011ml} and its discounting variant~\cite{czuba2024rank} are heuristics related to degree centrality. More recently, learning-based methods have also been developed to address the problem in such networks. For instance, GBIM~\cite{Yuan2024gbim} formulates multiplex influence maximisation as a graph Bayesian optimisation problem, employing an item-association network to govern the diffusion process. While conceptually appealing, the approach suffers from scalability issues in practice. Models like GNNRank~\cite{he2022gnnrank} and GraphTR~\cite{liu2020graph}, while potentially suitable for ranking tasks in heterogeneous graphs, suffer from similar transductive constraints. Unsupervised node embedding methods are also potentially applicable in this context, most notably Multi-node2vec~\cite{wilson2021analysis}, which learns continuous actor representations in multilayer networks. These embeddings can subsequently be clustered to estimate influence scores based on structural proximity. However, such two-stage pipelines introduce additional hyperparameter sensitivity and may lack end-to-end optimisation.

\paragraph{Conclusions.}

A review of the literature reveals that multilayer networks remain underexplored compared to classic graphs in the context of spreading processes, especially influence maximisation techniques, particularly those grounded in machine learning paradigms. Taken together, these limitations highlight a pressing need for robust, generalisable, and inductive methods addressig these challenges.

\subsection{Assumptions and main contributions}

In this study, we employ GNNs to model latent diffusion dynamics in multilayer networks and develop an end-to-end framework for identifying super-spreaders. This task can be viewed as a variant of the influence maximisation problem, which seeks to identify those agents within a graph that, if used as a single element seed set, are able to disseminate information in the most effective manner --- either by reaching the largest portion of the network or delivering it in the shortest possible time. Super-spreaders play a role analogous to that of a playmaker in a football team: if such a player is removed or underperforms, the team’s structure becomes destabilised and its winning potential is immediately diminished. 

We formulate the problem inspired by learning-to-rank task, in which the model ranks actors according to their spreading potential score, based on four features: (i) the number of activations; (ii) the duration of the diffusion process; (iii) the peak number of activations; and (iv) the simulation step at which this peak occurs --- predicted for a setting when the actor serves as the sole seed of the diffusion. Since, to the best of our knowledge, no such dataset exists, we construct one by generating hundreds of synthetic networks and simulating diffusion under the Independent Cascade Model~\cite{goldenberg2001icm} adapted to multilayer networks, to capture the aforementioned vectors.

Next, we propose a novel approach for identifying super-spreaders in multilayer networks by introducing the \textit{TopSpreadersNetwork} (\textit{ts-net}) --- a robust GNN capable of generalising to previously unseen data, a feature rarely found among competitive trainable solutions. This ability stems from an architectural design that enables training on a diverse collection of networks, allowing the model to capture transferable structural patterns. We demonstrate the practical applicability and scalability of the proposed solution by predicting top-spreaders in real-world networks of varying sizes, with no constraints on the number of actors or layers. Furthermore, a comparison with both classical heuristics and state-of-the-art trainable models shows that \textit{ts-net} consistently outperforms competitive solutions. The source code is available at: \url{https://github.com/network-science-lab/infmax-trainer-icm-mln}, appendices are published as~\cite{czuba2025identifyingsuperspreadersmultilayer}.

In summary, this work contributes by introducing:
\begin{itemize}[topsep=0pt]
    \item an inductive GNN model for super-spreader identification in multilayer networks, \textit{TopSpreadersNetwork}, capable of processing networks with varying sizes of actors and relations; 
    \item the  \textit{TopSpreadersDataset} constructed from simulations on synthetic and real-world multilayer networks, capturing the spreading potentials of actors;
    \item alternative approaches for data transformations and embedding fusion that enhance robustness in the ranking prediction task from multirelational data.
\end{itemize}

\section{Problem definition}

Before presenting the proposed model and dataset, we first need to introduce a formal definition of the problem under consideration, as well as the data structure relevant to this study.

\subsection{Multilayer networks}

To express our understanding of multilayer networks, we adopt the notation introduced by~\cite{kivela2014multilayer}, as shown in Def.~\ref{def:multilayer_net}.

\begin{definition}[Multilayer network]\label{def:multilayer_net}
    A multilayer network can be described as a tuple $M = (A,L,V,E)$ comprising the following sets:
    \begin{itemize}[noitemsep, topsep=0pt, label=-]
        \item actors $A=\{a_1, a_2, \dots, a_{|A|}\}$,
        \item layers $L=\{l_1, l_2, \dots, l_{|L|}\}$, 
        \item nodes $V=\{v_1^{1}, v_1^{2}, \dots, v_2^{1}, v_2^{2}, \dots \}: V \subseteq A \times L$,
        \item edges $E=\{(v_1^{1}, v_2^{1}), \dots, (v_1^{2}, v_2^{2}), \dots\}: (v_1^{1}, v_2^{2}) \notin E \land (v_1^{1}, v_2^{1}) \equiv (v_2^{1}, v_1^{1})$.
    \end{itemize}
\end{definition}

In essence, a multilayer network comprises a collection of intraconnected single-layer networks, where each actor may be represented by at most one node per layer, i.e., $V \subseteq A \times L$. For example, the representation of actor $a_1$ in layer $l_2$ is given by the node $v_1^2$. The model disallows interlayer edges, meaning that connections can occur only between nodes within the same layer: $(v_1^1, v_2^2) \notin E$. Furthermore, to simplify the analysis, all edges are assumed to be undirected and unweighted, that is, $(v_1^1, v_2^1) \equiv (v_2^1, v_1^1)$. At this point, it is worth noting that the concept of multilayer networks partially overlaps with that of heterogeneous graphs~\cite{Shi2022HeterogeneousGraphs}, the latter allowing a richer representation of the actor set (e.g., distinguishing between different classes of nodes) and permitting interlayer connections.

\subsection{Top-spreaders identification}

Since the goal of this study is to find an effective and applicable solution to the problem of super-spreaders identification, we must first formulate our approach to this task. To begin with, let us define a model $\phi$ that produces a ranking of actors (Def.~\ref{def:ml-model}) and a way such a ranking is created to reflect their spreading power (Def.~\ref{def:ts-ranking}).

\begin{definition}[Ranking predictor]\label{def:ml-model}
    Let $\phi(M) \rightarrow \mathbf{\hat{R}}$ be a function that predicts a ranking $\mathbf{\hat{R}}$ of actors from a multilayer network $M$.
\end{definition}

\begin{definition}[Top-spreaders ranking]\label{def:ts-ranking}
    A top-spreaders ranking $R$ is built upon a function $score: a \rightarrow \mathbb{R}^+$ that maps an actor $a$ to a value reflecting its spreading potential, defined as:
    \begin{equation*}
        \mathbf{R} = [a_i, a_j, \dots, a_k]: score(a_i) \geq score(a_j) \geq \dots \geq score(a_k)
    \end{equation*}
\end{definition}

It is important to note that, by introducing ground-truth ranking $\mathbf{R}$ and basing subsequent parts of the work on this notion (especially the methods used to construct ground-truth rankings), we deviate from the classic influence maximisation problem. This departure is justified by the non-additivity of seed sets: the top-$k$ actors or nodes that are most effective at spreading information individually do not necessarily constitute the optimal $k$-element seed set~\cite{kempe2003maximizing}.

Another important remark concerns the comparison between the ground-truth $\mathbf{R}$ and the predicted ranking $\mathbf{\hat{R}}$. A natural approach is to employ standard ranking-quality metrics such as Average Precision@$k$, or even the Jaccard index@$k$. However, in the problem we aim to address, the primary focus lies on the accuracy of the topmost predictions, where the most influential spreaders are expected to appear. Consequently, such metrics may be overly strict in evaluating discrepancies at the top of the ranking. For example, if the predicted ranking places the third and fourth most influential actors in the top two positions, it will yield the same Jaccard index as predicting two completely irrelevant actors in those positions, despite the former being a much more acceptable error.

To address this limitation and ensure a fairer comparison, we introduce the notion of a cumulated spreading score for the top-$k$ spreaders, as formalised in Def.~\ref{def:sps-cum}:

\begin{definition}[Cumulated spreading score]\label{def:sps-cum}
    The cumulated score of the top-$k$ spreaders in a network $M$ is defined as:
    \begin{equation*}
        y(\mathbf{R}, k) = \sum_{a \in \mathbf{R}[:k]} score(a)
    \end{equation*}
    where:
    \begin{description}
        \item $\mathbf{R}[:k]$ - first $k$ elements of the top-spreaders ranking list.
    \end{description}
\end{definition}

Based on the above, we can assess the quality of a predicted ranking $\mathbf{\hat{R}}$ in relation to the ground truth $\mathbf{R}$ (Def.~\ref{def:sps-relcum}). Instead of evaluating the prediction solely by the ordering of $\mathbf{\hat{R}}$, we compare the spreading scores of the top-$k$ actors in $\mathbf{\hat{R}}$ with those of the top-$k$ actors in $\mathbf{R}$.

\begin{definition}[Relative cumulated spreading score]\label{def:sps-relcum}
    The relative cumulated score of a predicted top-$k$ actors from a ranking $\mathbf{\hat{R}}$ w.r.t. the ground-truth ranking $\mathbf{R}$ is given by:
    \begin{equation*}
        y_{rel}(\mathbf{R}, \mathbf{\hat{R}}, k) = \frac{y(\mathbf{\hat{R}}, k)}{y(\mathbf{R}, k)}, \hspace{1em} y_{rel}(\cdot) \in [0, 1]
    \end{equation*}
\end{definition}

Finally, we summarise the objective of this study in formal terms by introducing the optimisation task presented in Def.~\ref{def:problem}, which aims to minimise the difference between $\mathbf{R}$ and $\mathbf{\hat{R}}$.

\begin{definition}[Top-spreaders identification problem]\label{def:problem}
    The task of identifying the top-$k$ spreaders in a network $M$, given a ground-truth ranking $R$, can be formulated as finding a function $\phi$ satisfying the following criterion:
    \begin{equation*}
        \phi(M) \rightarrow \mathbf{\hat{R}}: \min |1 - y_{rel}(\mathbf{R}, \mathbf{\hat{R}}, k)|
    \end{equation*}
\end{definition}

\section{Data preparation}
\label{dataset}

To facilitate the task of identifying super-spreaders in multilayer graphs, we first constructed a ground-truth dataset. The process consisted of three stages: collecting the graphs, simulating diffusion, and deriving a spreading potential score from the recorded sub-metrics.

\subsection{Collecting the networks}

We began by compiling a set of real-world multilayer graphs that are widely used within the network science community. In addition, using the \texttt{multinet} library~\cite{magnani2021analysis}, we generated one hundred Erd\H{o}s-R\'{e}nyi~\cite{er-model} and one hundred Preferential Attachment~\cite{sf-model} networks, supplementing these with eight graphs previously used in our own research. Tab.~\ref{tab:networks} presents collected networks with their basic parameters shortlisted.

\begin{table}[ht!]
    \centering
    \caption{Networks comprising the \textit{TopSpreadersDataset}, with their key parameters: number of layers, actors, nodes, edges, and the average degree per actor. For artificial graphs, mean values are reported.}
    \label{tab:networks}
    \addtolength{\tabcolsep}{-0.2em}
    \begin{tabular}{lrrrrr}
    Network type & Layers & Actors & Nodes & Edges & Degree \\ \hline
    artificial-er & 3.52 & 558.19 & 1741.70 & 6684.00 & 24.13 \\
    artificial-pa & 3.52 & 574.51 & 1976.07 & 42636.53 & 122.10\\
    artificial-small & 2.75 & 1000.00 & 2750.00 & 6609.12 & 13.22\\ \hline
    arxiv~\cite{dedomenico2015arxiv} & 13 & 14065 & 26796 & 59026 & 8.39 \\
    aucs~\cite{rossi2015aucs} & 5 & 61 & 224 & 620 & 20.33 \\
    ckmp~\cite{coleman1957ckmp} & 3 & 241 & 674 & 1370 & 11.37 \\
    eu-trans~\cite{cardillo2013eutransportation} & 37 & 417 & 2034 & 3588 & 17.21 \\
    l2-course~\cite{paradowski2024l2_course} & 2 & 41 & 82 & 297 & 14.49 \\
    lazega~\cite{snijders2006lazega} & 3 & 71 & 212 & 1659 & 46.73 \\
    timik~\cite{jankowski2024timik} & 3 & 61702 & 102247 & 875191 & 28.37 \\
    \end{tabular}
\end{table}

\subsection{Simulating information spreading}

The second step involved simulating diffusion under the Independent Cascade Model (ICM)~\cite{goldenberg2001icm}, adapted to account for the multilayer structure (MICM). Each actor in the network was used as a single-element seed set. During each simulation, we recorded four metrics that capture the spreading potential of the initiating actor, following~\cite{stolarski2024identifying}. These form a spreading-potential vector (Def.~\ref{def:spv}).

\begin{definition}[Spreading-potential vector]
    \label{def:spv}
    A vector of spreading potentials can be defined for each actor from the network $M$ as follows:
    \begin{equation*}
        \mathbf{p} = [p_{ex}, p_{sl}, p_{pi}, p_{pl}]
    \end{equation*}
    where:
    \begin{description}
        \item $p_{ex}$ - total number of activated actors across the network,
        \item $p_{sl}$ - spreading duration, as a number of MICM iterations,
        \item $p_{pi}$ - maximum number of actors activated simultaneously,
        \item $p_{pl}$ - iteration at which the activation peak occurred.
    \end{description}
\end{definition}

Enhancement of the ICM to multilayer graphs has been carried out based on the same premise as the extension of its similarly popular counterpart --- Linear Threshold Model (LTM), proposed by~\cite{zhong2022mltm}. Following the authors, we assume that actors are the primary subjects of the diffusion process, while the nodes representing them across different layers play an auxiliary role. In the classic ICM, a single parameter $\pi$ controls the probability that an active actor activates its inactive neighbour, with each activation attempt occurring only once. We augment this model with an additional parameter $\delta$, referred to as the \emph{protocol function}, which governs how activation signals from multiple layers are aggregated to determine the overall state of an actor. When the protocol condition is satisfied, the actor (and all nodes representing it) becomes active; otherwise, the actor remains inactive, even if some of its layer-specific representations receive positive input. We consider two edge cases for $\delta$: $AND$, where an actor is activated only if it receives positive input in \emph{all} layers where it is represented; and $OR$, where activation in \emph{any} layer is sufficient.

In total, we performed simulations under forty distinct configurations of MICM, with each experiment on a particular diffusion setup repeated forty times to reduce the stochastic effects inherent in the propagation model. Configurations of the MICM were defined by the Cartesian product of the two protocol functions: $\delta \in \{AND, OR\}$, and twenty activation probabilities: $\pi \in \{0.05, 0.10, \ldots, 1.00\}$. It is worth noting that we adopted the cascade-based model, rather than LTM, as it allows diffusion to be initiated from a single actor --- an essential requirement for computing individual spreading potentials.

\subsection{Defining spreading-potential score}

To enable ranking of actors according to Def.~\ref{def:ts-ranking}, we define a spreading-potential score that transforms the output vector $\mathbf{p}$ obtained during simulation into a unified scalar value.

\begin{definition}[Spreading-potential score]
    \label{def:sps}
    A score function that maps vectors of spreading potentials to a value in the range $[0, 1]$ is defined as:
    \begin{equation*}
        sps(\tilde{\mathbf{p}}) = \frac{1}{2} \tilde{p_{ex}} + \frac{1}{6} (1-\tilde{p_{sl}}) + \frac{1}{6} \tilde{p_{pi}} + \frac{1}{6} (1-\tilde{p_{pl}})
    \end{equation*}
    where:
    \begin{description}
        \item $\tilde{\mathbf{p}}$ - the coordinate-wise normalised $\mathbf{p}$ by maximal value occuring across the network 
    \end{description}
\end{definition}

The rationale behind this formulation reflects our interpretation of what constitutes a super-spreader. We prioritise the number of exposed actors, as it is typically regarded as the most critical metric in works related to spreading phenomena. All the remaining metrics are weighted equally, as none of them is deemed dominant. Moreover, our intention is to reward actors who spread information rapidly. Hence, we minimise $\tilde{p}_{sl}$ and $\tilde{p}_{pl}$ to favour fast and early spreaders. These weights, however, could also be arranged differently depending on the task. For instance, if one is concerned with abrupt diffusion, greater weight may be assigned to the activation peak ($\tilde{p_{pi}}$).

\begin{figure}[ht]
    \centering
    \subfloat[]{
    \includegraphics[page=1, width=0.3\linewidth]{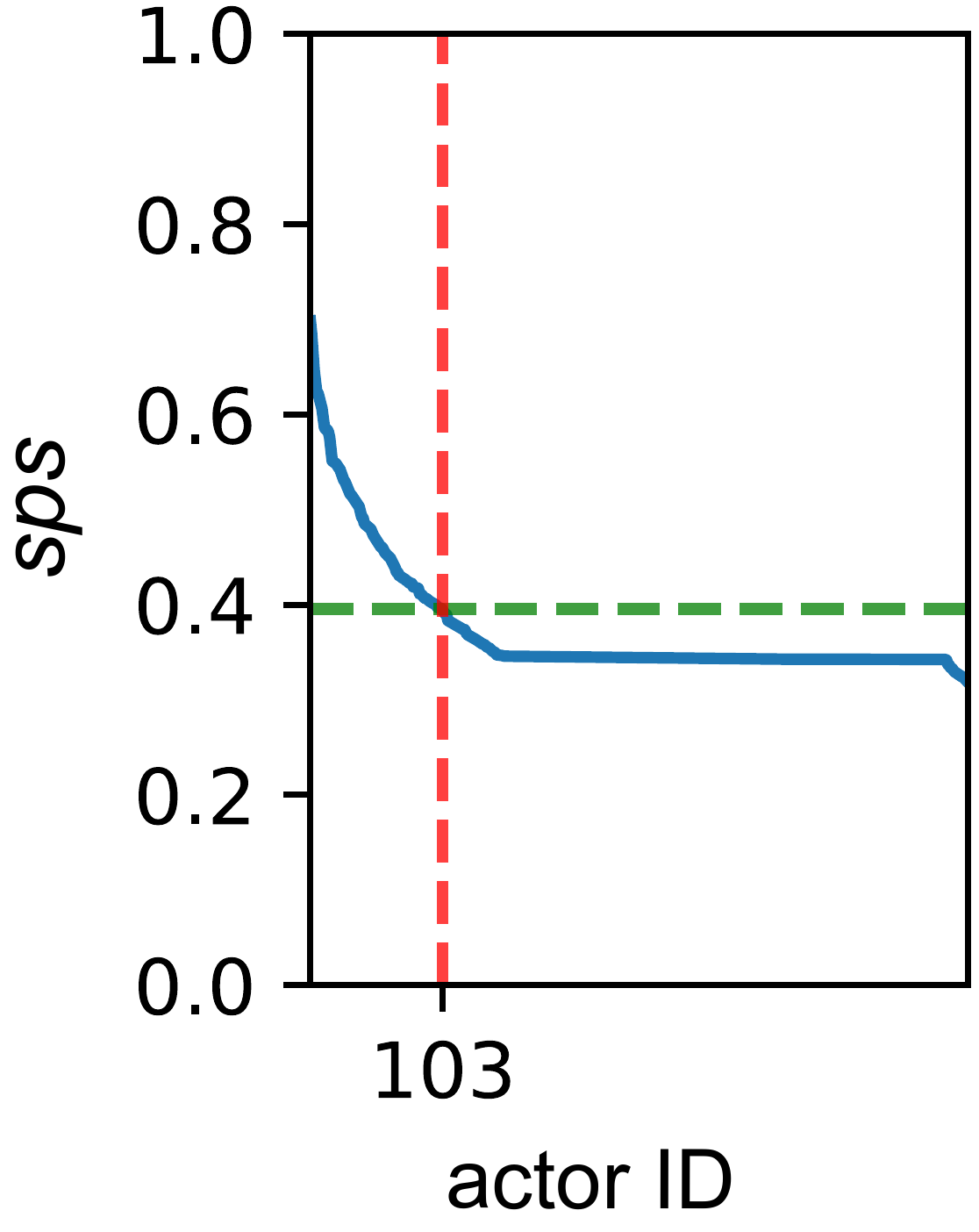}
    \label{fig:score_distribution_and}
    }
    \hspace{5em}
    \subfloat[]{
    \includegraphics[page=2, width=0.3\linewidth]{figures/distributions.pdf}
    \label{fig:score_distribution_or}
    }
    \caption{Distribution of $sps$ in network-72 from artificial-er split, obtained via simulations under $\delta=AND$ (\ref{fig:score_distribution_and}) and $\delta=OR$ (\ref{fig:score_distribution_or}). Dashed lines indicate the cutoff for the most significant spreaders.}
    \label{fig:score_distribution}
\end{figure}

Once the score values were computed for all networks, we identified feasible ranges of MICM parameters to be considered in subsequent experiments. This step was necessary due to the inherent behaviour of spreading models, where only a limited subset of parameters yields ``meaningful'' diffusion --- that is, processes that persist over several steps without either immediate extinction or full saturation. Finally, we determined these ranges separately for each protocol function, as their diffusion dynamics differ substantially. For $\delta = AND$, we selected $\pi \in \{0.80, 0.85, 0.90, 0.95\}$, while for $\delta = OR$, $\pi \in \{0.05, 0.10, 0.15, 0.20\}$. For each $\delta$, the final $sps$ value was obtained by averaging the vector $\mathbf{p}$ over the feasible values of $\pi$. Fig.~\ref{fig:score_distribution} presents a typical distribution of the resulting scores.

\subsection{Final metrics used in the study}
\label{subsec:metrics}

Following a detailed analysis of the $sps$ distributions, we observed two distinct patterns. Under the $\delta=AND$ protocol, the distribution typically exhibits a discernible point separating actors with high spreading potential from those with limited ability to propagate information. In contrast, the $\delta=OR$ protocol results in a flatter curve, with the majority of actors displaying similar spreading capability and a sharp decline in $sps$ values near the tail. Consequently, identifying top spreaders in the latter case becomes trivial, and we therefore excluded this data split from subsequent experiments.

This analysis led to the selection of the final metrics used to evaluate the performance of our proposed model and baseline methods. Primarily, we assessed the relative $sps$ of the actor identified as the most influential spreader in the network. Additionally, to account for the stratified nature of the $sps$ distribution, we defined a saddle ground-truth point $s$ at the 80th percentile for each network, enabling the evaluation of prediction quality for top spreaders. To assess the robustness of the methods, we also considered the area under the cumulative relative score curve, both up to the saddle point and over the entire range. For clarity, these metrics are summarised in Tab.~\ref{tab:metrics}. Finally, we note a technical detail: to ensure consistency, all distributions were normalised by the number of actors (i.e., $k \in \{1, \dots, |A|\}$ was transformed into $\tilde{k} \in \{\frac{1}{|A|}, \dots, 1\}$), resulting in relative prediction curves confined to the $[0, 1]$ range on both axes.

\begin{table}[ht]
    \caption{Final metrics, bounded within the range $[0, 1]$, used to assess the effectiveness of methods for identifying top spreaders.}
    \label{tab:metrics}
    \addtolength{\tabcolsep}{-0.2em}
    \renewcommand{\arraystretch}{2}
    \begin{tabular}{lcp{4cm}}
    Symbol & Formula & Description \\ \hline
    $T_{val}$ & $\displaystyle y_{rel}(\mathbf{R}, \mathbf{\hat{R}}, \frac{1}{|A|})$ & Relative score of the most significant spreader identified \\
    $S_{auc}$ & $\displaystyle \sum_{\frac{1}{|A|} \leq \tilde{k} \leq s} y_{rel}(\mathbf{R}, \mathbf{\hat{R}}, \tilde{k})$ & Area under the relative cumulated score curve up to the saddle point \\
    $S_{val}$ & $\displaystyle y_{rel}(\mathbf{R}, \mathbf{\hat{R}}, s)$ & Relative cumulated score of the identified top spreaders at the saddle point \\
    $F_{auc}$ & $\displaystyle \sum_{\frac{1}{|A|} \leq \tilde{k} \leq 1} \hspace{-.8em} y_{rel}(\mathbf{R}, \mathbf{\hat{R}}, \tilde{k})$ & Area under the full relative cumulated score curve
    \end{tabular}
\end{table}

\begin{figure*}[ht!]
    \centering
    \includegraphics[width=\linewidth]{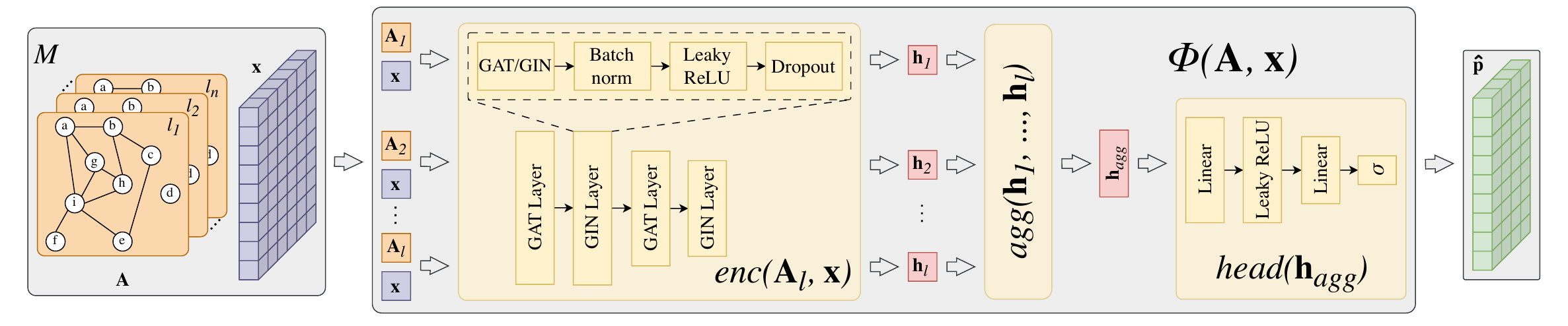}
    \caption{A schematic illustration of the \textit{ts-net} architecture processing a multilayer network. Each layer is encoded independently using a shared encoder composed of interleaved GAT and GIN blocks. The resulting representations are then aggregated by a trainable aggregation layer to produce actor embeddings. Finally, a vector of spreading potential is generated using a component consisting of multilayer perceptrons.}
    \label{fig:ts-net_architecture}
\end{figure*}

\section{Proposed model -- TopSpreadersNetwork}

One of the central contributions of this study is a supervised, inductive graph neural network model: \textit{ts-net}. The model comprises approximately $458{,}000$ trainable parameters and is designed to predict a vector of spreading potentials $\mathbf{\hat{p}}$, which is subsequently used to identify super-spreaders in size-adaptive, multilayer networks. The core concept of the model is illustrated in Fig.~\ref{fig:ts-net_architecture}. Owing to the adopted design choices, \textit{ts-net} produces robust predictions across networks of varying size, density, and structural complexity.

\subsection{Structural design}

The \textit{ts-net} is built upon three main components: (i) a shared encoder that produces actor embeddings for each layer of the network, (ii) a trainable aggregation module that combines these representations to derive final actor embeddings, and (iii) a prediction head that yields the output vector $\hat{\mathbf{p}}$ for each actor in the network. Each of these components is described in the following paragraphs.

\paragraph{Encoder.} The encoder is shared across all layers $L$ of the processed network $M$ and is responsible for computing actor embeddings independently for each layer $l \in L$. It processes the layer-specific adjacency matrix $\mathbf{A}$ and a feature vector $\mathbf{x}$. This module consists of a four-layer hybrid GNN architecture that combines GIN~\cite{xu2018powerful} and GAT~\cite{velivckovic2017graph} blocks, enhanced with dropout~\cite{hinton2012improving}, batch normalisation~\cite{ioffe2015batch}, and LeakyReLU~\cite{maas2013leakyrelu} activation. The hybrid design enables the model to capture rich structural context while maintaining strong generalisation capabilities, particularly in networks with complex topologies, aligning with the intuition presented in~\cite{pan2024hybridgnn} and supported by the experimental results (see App.~\ref{app:ablation_tuning}).

\begin{definition}[\textit{WiseAverage} fusion]
    \label{def:wise_fusion}
    The trainable aggregation mechanism that computes the actor embedding matrix $\mathbf{h}_{agg}$ from a collection of actor representations across network layers $\{\mathbf{h}_1, \dots, \mathbf{h}_{|L|}\}$, is defined as follows:
    \begin{align*}
    \mathbf{h}_{agg} &= \sum_{l=1}^{|L|} \mathbf{T}^{(l)} \odot \mathbf{h}_{l}, \quad 
    \mathbf{T} = \mathrm{softmax} \left(
    \begin{bmatrix}
    \mathbf{W} \mathbf{h}_{1} \\
    \mathbf{W} \mathbf{h}_{2} \\
    \vdots \\
    \mathbf{W} \mathbf{h}_{|L|}
    \end{bmatrix}
    \right)
    \end{align*}
    
    \begin{equation*}
    \begin{aligned}
        \mathbf{h}_{agg} &\in \mathbb{R}^{|A| \times d},
        \mathbf{T} \in \mathbb{R}^{|L| \times |A| \times 1},
        \mathbf{h}_l \in \mathbb{R}^{|A| \times d},
        \mathbf{W} \in \mathbb{R}^{1 \times d}
    \end{aligned}
    \end{equation*}
    where:
    \begin{description}
        \item $\mathbf{h}_{agg}$ - the final aggregated embedding matrix,
        \item $\mathbf{T}$ - the attention matrix computed via softmax,
        \item $\mathbf{h}_{l}$ - the embedding from the $l$-th layer of the processed network,
        \item $\mathbf{W}$ - a trainable weight matrix from the linear layer,
        \item $d$   - the embedding dimension,
        \item $|A|$ - the number of actors in the processed network,
        \item $|L|$ - the number of layers in the processed network.
    \end{description}
\end{definition}

\paragraph{Aggregation module.}

To combine actor representations from each layer of the processed network, we utilise a \textit{soft attention} mechanism~\cite{bahdanau2014neural}. Our approach --- \textit{WiseAverage} (Def.~\ref{def:wise_fusion}) is motivated by the intuition that an actor's relative importance may vary across layers; hence, aggregating layer-specific embeddings should be performed adaptively. Accordingly, each actor is assigned a normalised vector that reflects its importance on each layer and is then used to compute a weighted average of the layer-specific embeddings, yielding its network-level representation. As shown in the experimental part (see App.~\ref{app:ablation_tuning}), this fusion mechanism consistently outperforms several well-established embedding-level aggregation methods~\cite{bielak2024representation}.

\paragraph{Prediction head.}

The final component is built upon a two-layer MLP that predicts the vector $\mathbf{p}$ of spreading potentials (Def.~\ref{def:spv}), which is then used to compute scores and derive the ranking $\mathbf{\hat{R}}$.

\subsection{Techniques for robust prediction}

\begin{figure}[ht!]
    \centering
    \includegraphics[width=\linewidth]{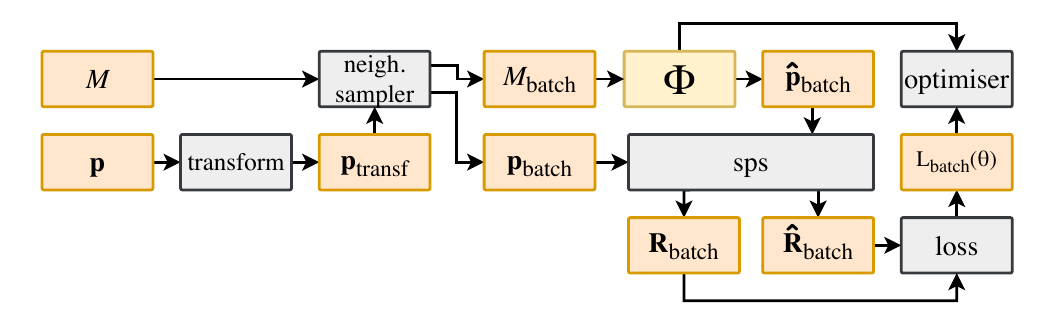}
    \caption{The pipeline used to train the \textit{ts-net}; orange blocks represent data entities, grey blocks indicate functional components, and the model is highlighted in yellow. Note that while the model is trained to optimise $\mathbf{\hat{R}}$, it is still capable of predicting $\mathbf{p}$.}
    \label{fig:training_pipeline}
\end{figure}

This section outlines the key design choices made to improve \textit{ts-net}'s predictive capabilities. Fig.~\ref{fig:training_pipeline} presents the data processing flow during training and inference. Although the former largely follows standard supervised learning, several non-trivial adaptations were introduced to improve model performance and generalisation.

\paragraph{Data transformations.}

Since the employed diffusion mechanism considers solely actor's connections across the network, no other actor attributes are meaningful in the context of this problem. Therefore, we use zero vectors as actor features, i.e., $\mathbf{x} = \mathbf{0} \in \mathbb{R}^{|A| \times d_{in}}$. As shown in App.~\ref{app:ablation_tuning}, this choice is supported by experiments involving alternative feature sets. Moreover, motivated by insights from the dataset analysis (Sec.~\ref{subsec:metrics}), we introduced a data transformation module into the training pipeline to enhance the visibility of super-spreaders. Specifically, we applied a function $transform: \mathbf{p} \rightarrow \mathbf{p}_{transf}$ to the ground-truth labels used during training. After evaluating the effectiveness of various approaches (see App.~\ref{app:ablation_tuning}), we selected one referred to as $scatter$, which amplifies the scores of high-potential spreaders while attenuating those of weaker ones. The transformation is defined as $\mathbf{p}_{transf} = \exp(3 \cdot \mathbf{p}) / \exp(3)$. Importantly, this function preserves interpretability, as it is invertible and allows for reconstruction of the original target values.

\paragraph{Local predictions.}

To maximise the throughput of the \textit{ts-net} when processing large and complex networks, we adopted the sampling technique introduced in GraphSAGE~\cite{hamilton2017inductive} (implemented in PyG~\cite{Fey2019torchgeometric} as the \texttt{NeighborLoader} class). The configuration was adjusted to ensure that only larger networks are subsampled, while smaller graphs are processed in full. This enables the model to learn from local rankings, significantly reducing computational requirements while allowing it to focus on the most relevant substructures within the network. As a result, prediction quality can be improved through better alignment with local topological patterns~\cite{chiang2019cluster}.

\paragraph{Optimisation task.}

For the training task, we employed the ListMLE loss function~\cite{xia2008listwise}, which is designed to optimise the model by sorting predicted scores. This enables the model to learn local rankings derived from neighbourhood samples. Importantly, although the model is trained to predict rankings $\mathbf{\hat{R}}$, it outputs vectors of spreading potentials $\mathbf{\hat{p}}$. This feature enhances the model’s interpretability, a characteristic often lacking in competitive methods, which typically do not provide explanations for their predictions.

\section{Evaluation and results}

A comprehensive set of experiments was designed to evaluate the effectiveness of \textit{ts-net} with respect to the following questions: (i) how does the proposed method perform on both synthetic and real-world multilayer networks compared to competitive solutions (ii) can an inductive approach achieve performance comparable to that of state-of-the-art approaches (iii) what are the limitations of the examined methods in handling structurally diverse graphs (iv) how do the individual components of \textit{ts-net} contribute to its overall performance (v) how do different hyperparameter settings, tasks, and dataset variants affect the performance of \textit{ts-net}? In the following section, we provide answers to most of these questions, while referring interested readers to App.~\ref{app:ablation_tuning} for a detailed discussion of the latter two.

\subsection{Experimental setup}
\label{sec:experiments}

To provide a comprehensive comparison of the capabilities of \textit{ts-net}, we selected seven competitive solutions, encompassing both heuristic methods and learning-based models. Notably, all of them address the broader problem of influence maximisation, rather than the more specific task of identifying top spreaders. Moreover, some were originally developed for homogeneous networks and thus required adaptation to the multilayer context. Our choice was intended to reflect a broad spectrum of successful, state-of-the-art methods that are most closely related in objective and scope to the considered problem.

The heuristic group includes classic centrality measures: Degree Centrality (\textit{deg-c}) and One-Hop Neighbourhood Size~\cite{magnani2011ml} (\textit{nghb-s}), along with their discounting counterparts: Degree Centrality Discount~\cite{chen2009efficient} (\textit{deg-cd}) and Neighbourhood Size Discount~\cite{czuba2024rank} (\textit{nghb-sd}). These were chosen for their low computational cost and ability to yield actor rankings compatible with multilayer settings. We also include a random baseline for reference. For learning-based models, we selected DeepIM~\cite{ling2023deepim} due to its data-driven design and recent impact in influence maximisation on homogeneous graphs. Despite its transductive nature, it serves as a strong neural baseline. As a representative of unsupervised embedding-based techniques adaptable to multilayer data, we selected Multi-node2vec~\cite{wilson2021analysis} with KMeans clustering (\textit{mn2v-km}) to represent unsupervised embedding-based techniques adaptable to multilayer data.

The experimental procedure consists of three stages. First, \textit{ts-net}, being the only inductive method among those evaluated, is trained on synthetic graphs, with $168$ used for training and $20$ for validation. In the second stage, all methods are applied to generate rankings of super spreaders on both synthetic and real-world networks not seen by the \textit{ts-net}. For transductive methods, training is performed individually on each test network at this stage. Finally, the resulting rankings are evaluated using the metrics listed in Tab.~\ref{tab:metrics}. We refer the reader to App.~\ref{app:eval_other_metrics} for results based on standard ranking metrics, which, however, are not substantially different from those presented below.

\subsection{Result analysis}

\begin{figure*}[ht!]
    \centering
    \subfloat[]{
    \includegraphics[page=1, width=0.43\linewidth]{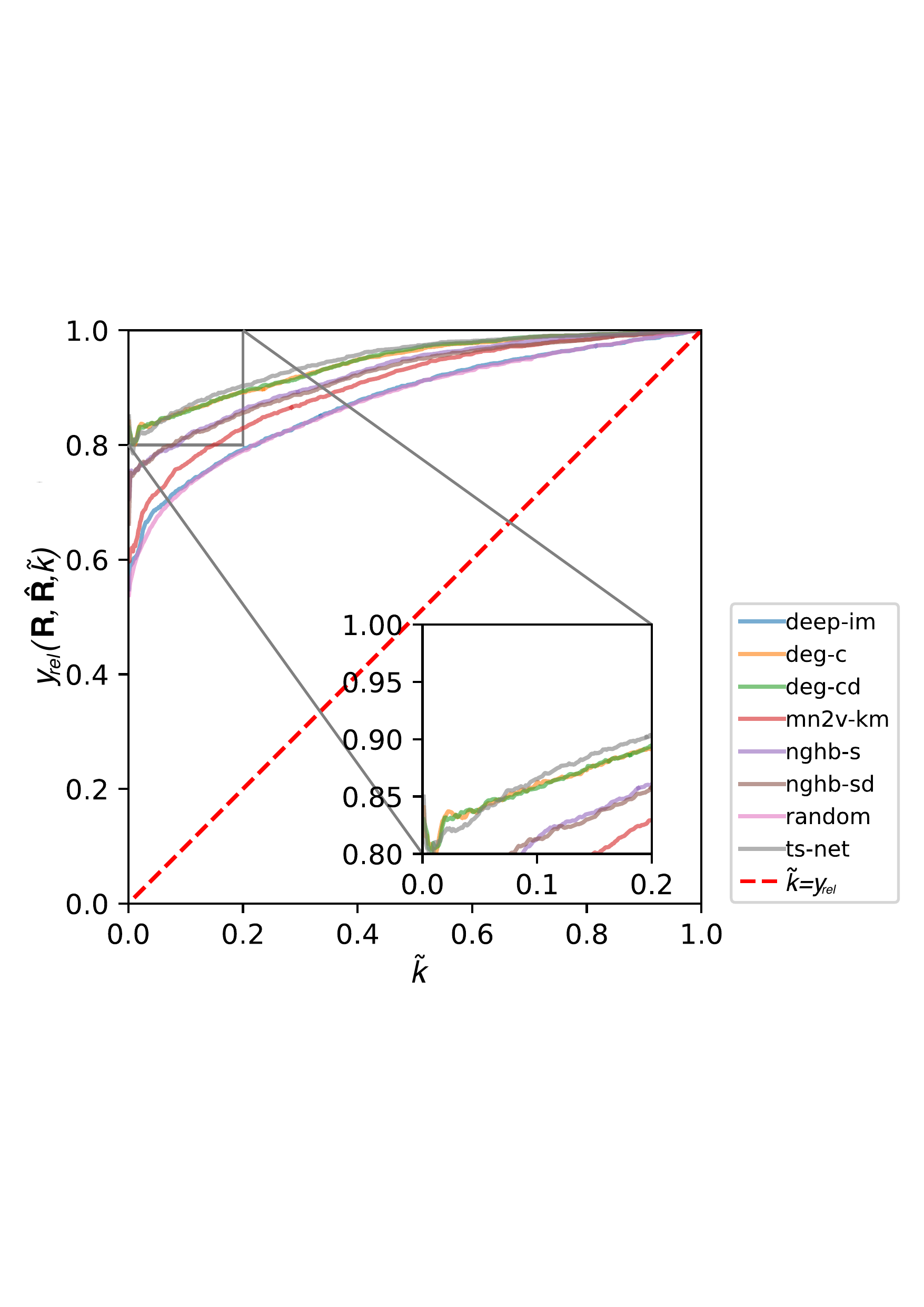}
    \label{fig:avg_curves_artificial}
    }
    \hspace{2em}
    \subfloat[]{
    \includegraphics[page=1, width=0.43\linewidth]{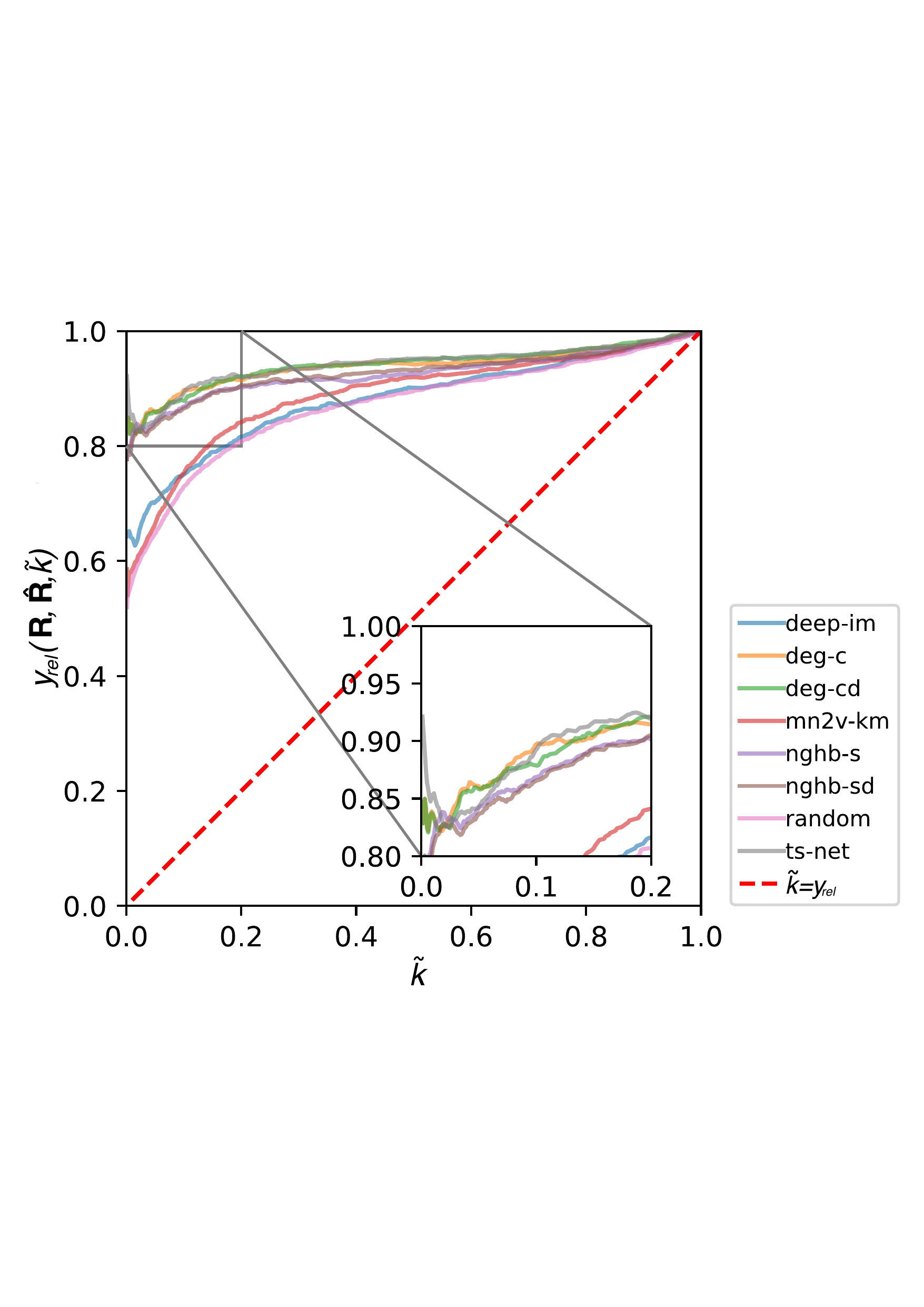}
    \label{fig:avg_curves_real}
    }
    \caption{Averaged cumulated relative score curves from prediction on test artificial (\ref{fig:avg_curves_artificial}) and real-world (\ref{fig:avg_curves_real}) networks by the evaluated methods.}
    \label{fig:avg_curves}
\end{figure*}

We begin the discussion of results by analysing the average performance of the evaluated methods, as illustrated by the cumulative relative score curves in Fig.~\ref{fig:avg_curves} and the corresponding metrics in Tab.~\ref{tab:avg_results}. \textit{ts-net} achieved the best results in $7$ out of $8$ scenarios, significantly outperforming all other methods, particularly on real-world networks under the $T_{val}$ metric, which captures the ability to identify the most influential spreader for the entire graph. The only case in which \textit{ts-net} was not ranked first was for real networks under the $S_{val}$ metric, where \textit{deg-cd} held a slight advantage (by $0.004$). Among the baselines, only \textit{deg-cd} and \textit{deg-c} delivered comparable results, while all other methods exhibited substantially lower performance. These findings are further supported by Fig.~\ref{fig:avg_curves}, which also reveals a marked drop in performance across most methods beyond the top-ranked actors, indicating room for further improvement and investigation.

\begin{table}[ht!]
    \centering
    \caption{Average performance of the evaluated methods across two network splits. Predictions with the highest quality (i.e., closest to~$1$) for each criterion are highlighted in green, with darker shades indicating better ranks.}
    \label{tab:avg_results}
    \addtolength{\tabcolsep}{-0.25em}
    \begin{tabular}{l|rrrr|rrrr}
    \multirow{2}{*}{$\phi$} & \multicolumn{4}{c|}{Artificial networks} & \multicolumn{4}{c}{Real networks} \\
        & $T_{val}$ & $S_{auc}$ & $S_{val}$ & $F_{auc}$ & $T_{val}$ & $S_{auc}$ & $S_{val}$ & $F_{auc}$ \\ \hline
    \textit{random} & 0.538 & 0.672 & 0.759 & 0.875 & 0.520 & 0.629 & 0.749 & 0.863 \\ \hline
    \textit{deg-c} & \second{0.841} & \second{0.821} & \second{0.872} & \third{0.942} & \second{0.829} & \second{0.819} & \third{0.896} & \third{0.931} \\
    \textit{deg-cd} & \third{0.830} & \third{0.820} & \second{0.872} & \second{0.943} & \second{0.829} & \third{0.815} & \first{0.901} & \second{0.933} \\
    \textit{nghb-s} & 0.706 & 0.772 & 0.834 & 0.924 & 0.777 & 0.806 & 0.887 & 0.918 \\
    \textit{nghb-sd} & 0.662 & 0.771 & 0.830 & 0.920 & 0.777 & 0.801 & 0.886 & 0.919 \\ \hline
    \textit{deep-im} & 0.548 & 0.679 & 0.761 & 0.878 & 0.643 & 0.677 & 0.788 & 0.872 \\
    \textit{mn2v-km} & 0.611 & 0.711 & 0.790 & 0.903 & 0.586 & 0.658 & 0.786 & 0.881 \\
    \textbf{\textit{ts-net}} & \first{0.850} & \first{0.826} & \first{0.881} & \first{0.947} & \first{0.921} & \first{0.826} & \second{0.897} & \first{0.934} \\
    \end{tabular}
\end{table}

A crucial aspect of GNNs is their ability to generalise. To examine this property, a detailed analysis of the performance on real-world networks, not seen by \textit{ts-net} during training, is presented in Tab.~\ref{tab:detailed_results}.

\begin{table*}[ht!]
    \centering
    \caption{Performance of the evaluated methods on the real-world networks. Predictions with the highest quality (i.e., closest to~$1$) for each criterion are highlighted in green, with darker shades indicating better ranks. $S_{auc}$ covers mostly with $S_{val}$ and is hidden for conciseness.}
    \label{tab:detailed_results}
    \addtolength{\tabcolsep}{-0.45em}
    \begin{tabular}{l|rrr|rrr|rrr|rrr|rrr|rrr|rrr}
     \multirow{2}{*}{$\phi$} & \multicolumn{3}{c|}{arxiv} & \multicolumn{3}{c|}{aucs} & \multicolumn{3}{c|}{ckmp} & \multicolumn{3}{c|}{eu-trans} & \multicolumn{3}{c|}{l2-course} & \multicolumn{3}{c|}{lazega} & \multicolumn{3}{c}{timik} \\
    & $T_{val}$ & $S_{val}$ & $F_{auc}$ & $T_{val}$ & $S_{val}$ & $F_{auc}$ & $T_{val}$ & $S_{val}$ & $F_{auc}$ & $T_{val}$ & $S_{val}$ & $F_{auc}$ & $T_{val}$ & $S_{val}$ & $F_{auc}$ & $T_{val}$ & $S_{val}$ & $F_{auc}$  & $T_{val}$ & $S_{val}$ & $F_{auc}$ \\ \hline
    \textit{random} & 0.390 & 0.810 & 0.920 & 0.640 & 0.810 & 0.876 & 0.460 & 0.790 & 0.878 & 0.420 & 0.750 & 0.887 & 0.740 & 0.840 & 0.859 & 0.660 & 0.750 & 0.809 & \third{0.330} & 0.490 & 0.813 \\ \hline
    \textit{deg-c} & \first{1.000} & \second{0.930} & \second{0.956} & \second{0.990} & \first{0.940} & \second{0.953} & \first{0.980} & 0.830 & 0.922 & \second{0.710} & 0.890 & 0.936 & \second{0.870} & \first{0.930} & \second{0.920} & \first{0.950} & \third{0.930} & \third{0.934} & 0.300 & \second{0.820} & \third{0.892} \\
    \textit{deg-cd} & \first{1.000} & 0.920 & 0.953 & \second{0.990} & \first{0.940} & \first{0.957} & \first{0.980} & \second{0.860} & \third{0.923} & \second{0.710} & \first{0.900} & \first{0.944} & \second{0.870} & \third{0.920} & \third{0.916} & \first{0.950} & \first{0.950} & \first{0.941} & 0.300 & \second{0.820} & \second{0.894} \\
    \textit{nghb-s} & \first{1.000} & \first{0.940} & \first{0.961} & 0.640 & 0.920 & 0.935 & \first{0.980} & 0.810 & 0.901 & 0.700 & \first{0.900} & \third{0.937} & \second{0.870} & 0.890 & 0.891 & \first{0.950} & \third{0.930} & 0.910 & 0.300 & \second{0.820} & 0.889 \\
    \textit{nghb-sd} & \first{1.000} & \second{0.930} & \third{0.955} & 0.640 & 0.930 & 0.939 & \first{0.980} & 0.810 & 0.905 & 0.700 & \first{0.900} & \second{0.943} & \second{0.870} & 0.890 & 0.893 & \first{0.950} & \second{0.940} & 0.911 & 0.300 & 0.800 & 0.888 \\ \hline
    \textit{deep-im} & 0.400 & 0.810 & 0.921 & 0.570 & 0.760 & 0.855 & 0.460 & 0.800 & 0.885 & 0.680 & 0.760 & 0.895 & \first{0.880} & 0.800 & 0.849 & 0.870 & 0.800 & 0.829 & OOM & OOM & OOM \\
    \textit{mn2v-km} & 0.520 & 0.820 & 0.924 & 0.550 & 0.770 & 0.866 & 0.480 & \first{0.880} & \first{0.927} & 0.400 & 0.740 & 0.882 & 0.800 & 0.820 & 0.842 & 0.390 & 0.870 & 0.859 & \second{0.960} & 0.600 & 0.868 \\
    \textbf{\textit{ts-net}} & 0.880 & 0.890 & 0.952 & \first{1.000} & \first{0.940} & \third{0.947} & \first{0.980} & \second{0.860} & \first{0.927} & \first{0.790} & 0.880 & 0.936 & \second{0.870} & \first{0.930} & \first{0.921} & \first{0.950} & \third{0.930} & \second{0.938} & \first{0.980} & \first{0.850} & \first{0.917} \\
    \end{tabular}
\end{table*}

Despite this setup, \textit{ts-net} was the only method to consistently achieve scores no lower than $0.790$ across all metrics and datasets. Moreover, it attained the best performance in $11$ out of the $21$ evaluation cases, outperforming all other methods; the second-best method, \textit{deg-cd}, was ranked first in $9$ cases. \textit{ts-net} obtained the highest $T_{val}$ score on five out of eight networks and consistently ranked among the top three in the remaining metrics. Its performance was especially pronounced on \textit{timik}, demonstrating that predicting local rankings to improve memory efficiency does not compromise predictive quality. Notable performance deviations were observed on the \textit{arxiv} and \textit{eu-trans} networks, both having considerably more layers ($13$ and $37$, respectively) than the maximum of $5$ in the training data. The performance drop on out-of-distribution networks may stem from limitations of the aggregation component. While scalability w.r.t. the number of actors has been achieved using \texttt{NeighborLoader}, the number of relationships requires further investigation. The observed effect similar to that of oversquashing (too much information from many layers is compressed into one vector) could be mitigated using techniques analogous to those employed for handling large neighbourhoods; i.e., by processing the layers ``chunk by chunk'', rather than all at once. Naturally, enriching the dataset with networks containing more layers may also help.

Among the other methods, \textit{deg-cd} and \textit{deg-c} yielded stable and competitive performance, especially on smaller networks, where they frequently ranked in the top three. However, they underperformed on more irregular graphs such as \textit{timik}. The results of \textit{nghb-s} and \textit{nghb-sd} were mixed: while effective in $T_{val}$ on \textit{arxiv}, \textit{ckmp}, and \textit{lazega}, their scores in $S_{val}$ and $F_{auc}$ were inconsistent. The \textit{random}, predictably, placed last in most cases. \textit{mn2v-km} performed reasonably on \textit{ckmp} and \textit{timik}, but fell behind top methods overall. Finally, \textit{DeepIM} performed poorly, except for $T_{val}$ on the smallest \textit{l2-course}, and failed to run on the largest graph (\textit{timik}) due to memory issues.

\section{Conclusions}

This study addresses the problem of identifying super-spreaders in multilayer networks by proposing a novel solution --- \textit{TopSpreadersNetwork} --- an inductive graph neural network capable of processing multilayer networks of varying sizes and complexities. To this end, we also constructed the \textit{TopSpreadersDataset}, which captures actor-level diffusion characteristics under the Multilayer Independent Cascade Model. Finally, we proposed and validated alternative techniques for data transformation and embedding fusion that improve the robustness of ranking-based learning from multirelational data.

Nonetheless, some limitations remain. While multilayer networks constitute a powerful and flexible framework, they are still a subset of the broader class of heterogeneous graphs, which are not directly addressed in this work. Moreover, although \textit{TopSpreadersNetwork} demonstrates superior performance, relatively simple heuristics such as Degree Centrality continue to yield strong results (for a comparison of time performance see App.~\ref{app:time_complexity}). The main advantage of our approach lies in its interpretability: rather than producing scores in a black-box fashion, it first predicts a vector of four interpretable features, which can then be used to construct tailored rankings. This offers flexibility for use cases where, e.g, actors causing the highest peak of infections are more desirable than those triggering longer diffusion, something other solutions do not easily allow.

Several directions exist for future research. One interesting option is to include diffusion results for small yet effective seed sets, e.g., generated using the greedy algorithm, thereby extending the learning task from scoring individual actors to evaluating entire sets. Another promising avenue is to reforge the dataset into a general benchmark for super-spreader identification, encompassing a broader range of diffusion models and network topologies. 



\begin{ack}
This research was partially supported by the National Science Centre, Poland, grant no. 2022/45/B/ST6/04145, the Polish Ministry of Science and Higher Education programme: International Projects Co-Funded, and the EU under the Horizon Europe, grant no. 101086321. Views and opinions expressed here are those of the authors only and do not necessarily reflect those of the funding agencies.
\end{ack}


\clearpage
\bibliography{references}

\begin{thebibliography}{51}
\providecommand{\natexlab}[1]{#1}
\providecommand{\url}[1]{\texttt{#1}}
\expandafter\ifx\csname urlstyle\endcsname\relax
  \providecommand{\doi}[1]{doi: #1}\else
  \providecommand{\doi}{doi: \begingroup \urlstyle{rm}\Url}\fi

\bibitem[Bahdanau et~al.(2014)Bahdanau, Cho, and Bengio]{bahdanau2014neural}
D.~Bahdanau, K.~Cho, and Y.~Bengio.
\newblock Neural machine translation by jointly learning to align and translate.
\newblock \emph{arXiv preprint:1409.0473}, 2014.

\bibitem[Bielak and Kajdanowicz(2024)]{bielak2024representation}
P.~Bielak and T.~Kajdanowicz.
\newblock Representation learning in multiplex graphs: Where and how to fuse information?
\newblock In \emph{International Conference on Computational Science}, pages 3--18. Springer, 2024.

\bibitem[Bollobas et~al.(2003)Bollobas, Borgs, Chayes, and Riordan]{sf-model}
B.~Bollobas, C.~Borgs, J.~Chayes, and O.~Riordan.
\newblock Directed scale-free graphs.
\newblock In \emph{Proceedings of the 14th Annual ACM-SIAM Symposium on Discrete Algorithms (SODA)}, pages 132--139, 1 2003.

\bibitem[Cardillo et~al.(2013)Cardillo, G{\ifmmode\acute{o}\else\'{o}\fi}mez-Garde{\ifmmode\tilde{n}\else\~{n}\fi}es, Zanin, Romance, Papo, Pozo, and Boccaletti]{cardillo2013eutransportation}
A.~Cardillo, J.~G{\ifmmode\acute{o}\else\'{o}\fi}mez-Garde{\ifmmode\tilde{n}\else\~{n}\fi}es, M.~Zanin, M.~Romance, D.~Papo, F.~d. Pozo, and S.~Boccaletti.
\newblock {Emergence of network features from multiplexity}.
\newblock \emph{Scientific Reports}, 3\penalty0 (1344):\penalty0 1--6, 2 2013.
\newblock ISSN 2045-2322.

\bibitem[Chamberlain et~al.(2021)Chamberlain, Rowbottom, Gorinova, Bronstein, Webb, and Rossi]{diff1}
B.~Chamberlain, J.~Rowbottom, M.~I. Gorinova, M.~Bronstein, S.~Webb, and E.~Rossi.
\newblock Grand: Graph neural diffusion.
\newblock In M.~Meila and T.~Zhang, editors, \emph{Proceedings of the 38th International Conference on Machine Learning}, volume 139 of \emph{Proceedings of Machine Learning Research}, pages 1407--1418. PMLR, 18--24 Jul 2021.

\bibitem[Chen et~al.(2009)Chen, Wang, and Yang]{chen2009efficient}
W.~Chen, Y.~Wang, and S.~Yang.
\newblock Efficient influence maximization in social networks.
\newblock In \emph{Proceedings of the 15th ACM SIGKDD internat. conf. on Knowledge discovery and data mining}, pages 199--208, 2009.

\bibitem[Chiang et~al.(2019)Chiang, Liu, Si, Li, Bengio, and Hsieh]{chiang2019cluster}
W.-L. Chiang, X.~Liu, S.~Si, Y.~Li, S.~Bengio, and C.-J. Hsieh.
\newblock Cluster-gcn: An efficient algorithm for training deep and large graph convolutional networks.
\newblock In \emph{Proceedings of the 25th ACM SIGKDD internat. conf. on knowledge discovery and data mining}, pages 257--266, 2019.

\bibitem[Coleman et~al.(1957)Coleman, Katz, and Menzel]{coleman1957ckmp}
J.~Coleman, E.~Katz, and H.~Menzel.
\newblock The diffusion of an innovation among physicians.
\newblock \emph{Sociometry}, 20\penalty0 (4):\penalty0 253--270, 1957.
\newblock ISSN 00380431.

\bibitem[Czuba and Br{\'o}dka(2025)]{czuba2024rank}
M.~Czuba and P.~Br{\'o}dka.
\newblock {Rank-refining seed selection methods for budget constrained influence maximisation in multilayer networks under linear threshold model}.
\newblock \emph{Social Network Analysis and Mining}, 15\penalty0 (1):\penalty0 46--25, Apr. 2025.
\newblock ISSN 1869-5469.

\bibitem[Czuba et~al.(2025)Czuba, Stolarski, Piróg, Bielak, and Bródka]{czuba2025identifyingsuperspreadersmultilayer}
M.~Czuba, M.~Stolarski, A.~Piróg, P.~Bielak, and P.~Bródka.
\newblock Identifying super spreaders in multilayer networks [preprint with appendices].
\newblock \emph{arXiv}, May 2025.
\newblock \doi{10.48550/arXiv.2505.20980}.

\bibitem[De~Domenico et~al.(2015)De~Domenico, Lancichinetti, Arenas, and Rosvall]{dedomenico2015arxiv}
M.~De~Domenico, A.~Lancichinetti, A.~Arenas, and M.~Rosvall.
\newblock Identifying modular flows on multilayer networks reveals highly overlapping organization in interconnected systems.
\newblock \emph{Phys. Rev. X}, 5:\penalty0 011027, 3 2015.

\bibitem[Deng et~al.(2020)Deng, Long, Li, Cao, and Pan]{Deng2020}
X.~Deng, F.~Long, B.~Li, D.~Cao, and Y.~Pan.
\newblock An influence model based on heterogeneous online social network for influence maximization.
\newblock \emph{IEEE Transactions on Network Science and Eng.}, 7:\penalty0 737--749, 2020.

\bibitem[Erd{\H{o}}s et~al.(1960)Erd{\H{o}}s, R{\'e}nyi, et~al.]{er-model}
P.~Erd{\H{o}}s, A.~R{\'e}nyi, et~al.
\newblock On the evolution of random graphs.
\newblock \emph{Publ. math. inst. hung. acad. sci}, 5\penalty0 (1):\penalty0 17--60, 1960.

\bibitem[Fey and Lenssen(2019)]{Fey2019torchgeometric}
M.~Fey and J.~E. Lenssen.
\newblock Fast graph representation learning with {PyTorch Geometric}.
\newblock In \emph{ICLR Workshop on Representation Learning on Graphs and Manifolds}, 2019.

\bibitem[Goldenberg et~al.(2001)Goldenberg, Libai, and Muller]{goldenberg2001icm}
J.~Goldenberg, B.~Libai, and E.~Muller.
\newblock {Talk of the Network: A Complex Systems Look at the Underlying Process of Word-of-Mouth}.
\newblock \emph{Marketing Letters}, 12\penalty0 (3):\penalty0 211--223, Aug. 2001.
\newblock ISSN 1573-059X.

\bibitem[Hamilton et~al.(2017)Hamilton, Ying, and Leskovec]{hamilton2017inductive}
W.~Hamilton, Z.~Ying, and J.~Leskovec.
\newblock Inductive representation learning on large graphs.
\newblock \emph{Advances in neural inf. processing sys.}, 30, 2017.

\bibitem[He et~al.(2022)He, Gan, Wipf, Reinert, Yan, and Cucuringu]{he2022gnnrank}
Y.~He, Q.~Gan, D.~Wipf, G.~D. Reinert, J.~Yan, and M.~Cucuringu.
\newblock Gnnrank: Learning global rankings from pairwise comparisons via directed gnns.
\newblock In \emph{international conference on machine learning}, 2022.

\bibitem[Hinton et~al.(2012)Hinton, Srivastava, Krizhevsky, Sutskever, and Salakhutdinov]{hinton2012improving}
G.~E. Hinton, N.~Srivastava, A.~Krizhevsky, I.~Sutskever, and R.~R. Salakhutdinov.
\newblock Improving neural networks by preventing co-adaptation of feature detectors.
\newblock \emph{arXiv preprint arXiv:1207.0580}, 2012.

\bibitem[Ioffe and Szegedy(2015)]{ioffe2015batch}
S.~Ioffe and C.~Szegedy.
\newblock Batch normalization: Accelerating deep network training by reducing internal covariate shift.
\newblock In \emph{International conference on machine learning}, pages 448--456. pmlr, 2015.

\bibitem[Jankowski et~al.(2017)Jankowski, Michalski, and Br\'{o}dka]{jankowski2024timik}
J.~Jankowski, R.~Michalski, and P.~Br\'{o}dka.
\newblock A multilayer network dataset of interaction and influence spreading in a virtual world.
\newblock \emph{Scientific Data}, 4\penalty0 (1):\penalty0 170144, 2017.
\newblock ISSN 2052-4463.

\bibitem[K et~al.(2022)K, Katukuri, and Jagarapu]{Venkatakrishna2022CIM}
V.~{\relax Rao}. K, M.~Katukuri, and M.~Jagarapu.
\newblock {CIM: clique-based heuristic for finding influential nodes in multilayer networks}.
\newblock \emph{Applied Intelligence}, 52\penalty0 (5):\penalty0 5173--5184, Mar. 2022.
\newblock ISSN 1573-7497.

\bibitem[Kempe et~al.(2003)Kempe, Kleinberg, and Tardos]{kempe2003maximizing}
D.~Kempe, J.~Kleinberg, and {\'E}.~Tardos.
\newblock Maximizing the spread of influence through a social network.
\newblock In \emph{Proceedings of the 9th ACM SIGKDD international conference on Knowledge discovery and data mining}, pages 137--146, 2003.

\bibitem[Kipf and Welling(2017)]{gcn}
T.~N. Kipf and M.~Welling.
\newblock Semi-supervised classification with graph convolutional networks, 2017.

\bibitem[Kivel{\"a} et~al.(2014)Kivel{\"a}, Arenas, Barthelemy, Gleeson, Moreno, and Porter]{kivela2014multilayer}
M.~Kivel{\"a}, A.~Arenas, M.~Barthelemy, J.~P. Gleeson, Y.~Moreno, and M.~A. Porter.
\newblock Multilayer networks.
\newblock \emph{Journal of Complex Networks}, 2\penalty0 (3):\penalty0 203--271, 07 2014.
\newblock ISSN 2051-1310.

\bibitem[Kumar et~al.(2022)Kumar, Mallik, Khetarpal, and Panda]{kumar2022}
S.~Kumar, A.~Mallik, A.~Khetarpal, and B.~Panda.
\newblock Influence maximization in social networks using graph embedding and graph neural network.
\newblock \emph{Information Sciences}, 607:\penalty0 1617--1636, 2022.
\newblock ISSN 0020-0255.

\bibitem[Li et~al.(2022)Li, Li, Liu, and Li]{mahe2022b}
Y.~Li, L.~Li, Y.~Liu, and Q.~Li.
\newblock Mahe-im: Multiple aggregation of heterogeneous relation embedding for influence max. on heterogeneous information network.
\newblock \emph{Expert Sys. with Appl.}, 202:\penalty0 117289, 04 2022.

\bibitem[Ling et~al.(2022{\natexlab{a}})Ling, Chowdhury, Jiang, Wang, Zhang, Chen, and Zhao]{graphreasoning}
C.~Ling, T.~Chowdhury, J.~Jiang, J.~Wang, X.~Zhang, H.~Chen, and L.~Zhao.
\newblock Deepgar: Deep graph learning for analogical reasoning, 2022{\natexlab{a}}.

\bibitem[Ling et~al.(2022{\natexlab{b}})Ling, Jiang, Wang, and Liang]{diff_source2}
C.~Ling, J.~Jiang, J.~Wang, and Z.~Liang.
\newblock Source localization of graph diffusion via variational autoencoders for graph inverse problems.
\newblock In \emph{Proceedings of the 28th ACM SIGKDD Conference on Knowledge Discovery and Data Mining}, KDD ’22, page 1010–1020. ACM, Aug. 2022{\natexlab{b}}.

\bibitem[Ling et~al.(2023{\natexlab{a}})Ling, Cao, and Zhao]{graphgeneration}
C.~Ling, H.~Cao, and L.~Zhao.
\newblock {STGEN: Deep Continuous-Time Spatiotemporal Graph Generation}.
\newblock In \emph{{Machine Learning and Knowledge Discovery in Databases}}, pages 340--356. Springer, Cham, Switzerland, Mar. 2023{\natexlab{a}}.
\newblock ISBN 978-3-031-26409-2.

\bibitem[Ling et~al.(2023{\natexlab{b}})Ling, Jiang, Wang, Thai, Xue, Song, Qiu, and Zhao]{ling2023deepim}
C.~Ling, J.~Jiang, J.~Wang, M.~T. Thai, L.~Xue, J.~Song, M.~Qiu, and L.~Zhao.
\newblock Deep graph representation learning and optimization for influence maximization.
\newblock In \emph{Proceedings of the 40th International Conference on Machine Learning}, ICML'23. JMLR.org, 2023{\natexlab{b}}.
\newblock \doi{10.5555/3618408.3619288}.

\bibitem[Liu et~al.(2020)Liu, Xie, Chen, Liu, Tu, Cui, Zhang, and Lin]{liu2020graph}
Q.~Liu, R.~Xie, L.~Chen, S.~Liu, K.~Tu, P.~Cui, B.~Zhang, and L.~Lin.
\newblock Graph neural network for tag ranking in tag-enhanced video recommendation.
\newblock In \emph{Proceedings of the 29th ACM international conference on information \& knowledge management}, 2020.

\bibitem[Maas et~al.(2013)Maas, Hannun, and Ng]{maas2013leakyrelu}
A.~L. Maas, A.~Y. Hannun, and A.~Y. Ng.
\newblock Rectifier nonlinearities improve neural network acoustic models.
\newblock In \emph{Proceedings of the 30th International Conference on Machine Learning (ICML)}, 2013.
\newblock ICML Workshop on Deep Learning for Audio, Speech and Language Processing.

\bibitem[Magnani and Rossi(2011)]{magnani2011ml}
M.~Magnani and L.~Rossi.
\newblock The ml-model for multi-layer social networks.
\newblock In \emph{2011 International conference on advances in social networks analysis and mining}, pages 5--12. IEEE, 2011.

\bibitem[Magnani et~al.(2021)Magnani, Rossi, and Vega]{magnani2021analysis}
M.~Magnani, L.~Rossi, and D.~Vega.
\newblock Analysis of multiplex social networks with r.
\newblock \emph{Journal of Statistical Software}, 98\penalty0 (8):\penalty0 1–30, 2021.

\bibitem[Pan et~al.(2024)Pan, Qu, Yao, Wang, et~al.]{pan2024hybridgnn}
C.-H. Pan, Y.~Qu, Y.~Yao, M.-J.-S. Wang, et~al.
\newblock Hybridgnn: A self-supervised graph neural network for efficient maximum matching in bipartite graphs.
\newblock \emph{Symmetry}, 16\penalty0 (12):\penalty0 1631, 2024.

\bibitem[Paradowski et~al.(2024)Paradowski, Whitby, Czuba, and Br{\ifmmode\acute{o}\else\'{o}\fi}dka]{paradowski2024l2_course}
M.~B. Paradowski, N.~Whitby, M.~Czuba, and P.~Br{\ifmmode\acute{o}\else\'{o}\fi}dka.
\newblock {Peer Interaction Dynamics and L2 Learning Trajectories During Study Abroad: A Longitudinal Investigation Using Dynamic Computational Social Network Analysis}.
\newblock \emph{Language Learning}, 74\penalty0 (S2):\penalty0 58--115, Dec. 2024.

\bibitem[Rossi and Magnani(2015)]{rossi2015aucs}
L.~Rossi and M.~Magnani.
\newblock Towards effective visual analytics on multiplex and multilayer networks.
\newblock \emph{Chaos, Solitons \& Fractals}, 72:\penalty0 68--76, 2015.
\newblock ISSN 0960-0779.

\bibitem[Salehi et~al.(2015)Salehi, Sharma, Marzolla, Magnani, Siyari, and Montesi]{magnani2015spreadingmln}
M.~Salehi, R.~Sharma, M.~Marzolla, M.~Magnani, P.~Siyari, and D.~Montesi.
\newblock Spreading processes in multilayer networks.
\newblock \emph{IEEE Transactions on Network Science and Engineering}, 2\penalty0 (2):\penalty0 65--83, 2015.

\bibitem[Shi et~al.(2022)Shi, Wang, and Yu]{Shi2022HeterogeneousGraphs}
C.~Shi, X.~Wang, and P.~S. Yu.
\newblock \emph{Introduction}, pages 1--8.
\newblock Springer Singapore, Singapore, 2022.
\newblock ISBN 978-981-16-6166-2.

\bibitem[Singh et~al.(2022)Singh, Srivastva, Verma, and Singh]{singh2022influence}
S.~S. Singh, D.~Srivastva, M.~Verma, and J.~Singh.
\newblock Influence maximization frameworks, performance, challenges and directions on social network: A theoretical study.
\newblock \emph{Journal of King Saud University-Computer and Information Sciences}, 34\penalty0 (9):\penalty0 7570--7603, 2022.

\bibitem[Snijders et~al.(2006)Snijders, Pattison, Robins, and Handcock]{snijders2006lazega}
T.~A.~B. Snijders, P.~E. Pattison, G.~L. Robins, and M.~S. Handcock.
\newblock New specifications for exponential random graph models.
\newblock \emph{Sociological Methodology}, 36\penalty0 (1):\penalty0 99--153, 2006.

\bibitem[Stolarski et~al.(2024)Stolarski, Pir{\'o}g, and Br{\'o}dka]{stolarski2024identifying}
M.~Stolarski, A.~Pir{\'o}g, and P.~Br{\'o}dka.
\newblock Identifying key nodes for the influence spread using a ml approach.
\newblock \emph{Entropy}, 26\penalty0 (11):\penalty0 955, 2024.

\bibitem[Tian et~al.(2020)Tian, Mo, Wang, and Peng]{Tian2020topicAwareIM}
S.~Tian, S.~Mo, L.~Wang, and Z.~Peng.
\newblock {Deep Reinforcement Learning-Based Approach to Tackle Topic-Aware Influence Maximization}.
\newblock \emph{Data Science and Engineering}, 5\penalty0 (1):\penalty0 1--11, Mar. 2020.
\newblock ISSN 2364-1541.

\bibitem[Veli{\v{c}}kovi{\'c} et~al.(2017)Veli{\v{c}}kovi{\'c}, Cucurull, Casanova, Romero, Lio, and Bengio]{velivckovic2017graph}
P.~Veli{\v{c}}kovi{\'c}, G.~Cucurull, A.~Casanova, A.~Romero, P.~Lio, and Y.~Bengio.
\newblock Graph attention networks.
\newblock \emph{arXiv preprint arXiv:1710.10903}, 2017.

\bibitem[Wang et~al.(2022)Wang, Jiang, and Zhao]{diff_source1}
J.~Wang, J.~Jiang, and L.~Zhao.
\newblock An invertible graph diffusion neural network for source localization, 2022.

\bibitem[Wilson et~al.(2021)Wilson, Baybay, Sankar, Stillman, and Popa]{wilson2021analysis}
J.~D. Wilson, M.~Baybay, R.~Sankar, P.~Stillman, and A.~M. Popa.
\newblock Analysis of population functional connectivity data via multilayer network embeddings.
\newblock \emph{Network Science}, 9\penalty0 (1):\penalty0 99--122, 2021.

\bibitem[Wu et~al.(2021)Wu, Pan, Chen, Long, Zhang, and Yu]{gnnSurv}
Z.~Wu, S.~Pan, F.~Chen, G.~Long, C.~Zhang, and P.~S. Yu.
\newblock A comprehensive survey on graph neural networks.
\newblock \emph{IEEE Transactions on Neural Networks and Learning Systems}, 32\penalty0 (1):\penalty0 4--24, 2021.

\bibitem[Xia et~al.(2008)Xia, Liu, Wang, Zhang, and Li]{xia2008listwise}
F.~Xia, T.-Y. Liu, J.~Wang, W.~Zhang, and H.~Li.
\newblock Listwise approach to learning to rank: theory and algorithm.
\newblock In \emph{Proceedings of the 25th international conference on Machine learning}, pages 1192--1199, 2008.

\bibitem[Xu et~al.(2018)Xu, Hu, Leskovec, and Jegelka]{xu2018powerful}
K.~Xu, W.~Hu, J.~Leskovec, and S.~Jegelka.
\newblock How powerful are graph neural networks?
\newblock \emph{arXiv preprint arXiv:1810.00826}, 2018.

\bibitem[Yuan et~al.(2024)Yuan, Shao, and Chen]{Yuan2024gbim}
Z.~Yuan, M.~Shao, and Z.~Chen.
\newblock {Graph Bayesian Optimization for Multiplex Influence Maximization}.
\newblock In \emph{Proceedings of the AAAI Conference on Artificial Intelligence}, Mar. 2024.

\bibitem[Zhong et~al.(2022)Zhong, Srivastava, and Leonard]{zhong2022mltm}
Y.~D. Zhong, V.~Srivastava, and N.~E. Leonard.
\newblock Influence spread in the heterogeneous multiplex linear threshold model.
\newblock \emph{IEEE Transactions on Control of Network Systems}, 9\penalty0 (3):\penalty0 1080--1091, 2022.

\end{thebibliography}


\clearpage

\appendix

\section{Model selection and ablation study}
\label{app:ablation_tuning}

In this section, we present experiments on the \textit{ts-net}, exploring various architectural designs, training configurations, and dataset settings. As illustrated in Fig.~\ref{fig:training_pipeline}, the training pipeline comprises several components, each of which could be examined in detail. However, to maintain conciseness, we report only the most essential aspects.

In each of the following subsections, the baseline architecture of the \textit{ts-net} remains unchanged. Specifically, we employ an encoder composed of a sequence of GAT–GIN–GAT–GIN neural layers, with a maximum hidden dimensionality of 400 channels. Each layer is followed by batch normalisation, a LeakyReLU activation, and dropout. For the aggregation layer, we use the \textit{WiseAverage} ($avg_w$) block, whereas vector of zeros serves as a feature matrix $\mathbf{x}$ and the spreading-potential scores are normalised columnwise by a maximal value. This design was selected on the basis that it proved most effective in preliminary experiments. In each table, we then present how the model's performance is affected when a specific parameter of the model or the training pipeline is modified.

\subsection{Model}

We begin with an ablation study of the model itself. Following the data flow direction, we evaluate the architectural designs of the encoder layers (Tab.~\ref{tab:ablation_model_encoders}), the number of hidden channels in the encoder (Tab.~\ref{tab:ablation_model_channels}), and the aggregation layers (Tab.~\ref{tab:ablation_model_aggregations}).

\begin{table}[ht!]
    \centering
    \caption{Performance of \textit{ts-net} with encoder's architectures; $S_{auc}$ covers mostly with $S_{val}$ and is hidden for concisness.}
    \label{tab:ablation_model_encoders}
    \addtolength{\tabcolsep}{-0.5em}
    \begin{tabular}{ll|rrr|rrrr}
    \multirow{2}{*}{Neural layers}& \multirow{2}{*}{Norm. and activat.} & \multicolumn{3}{c|}{Artificial networks} & \multicolumn{3}{c}{Real networks} \\
     & & $T_{val}$ &  $S_{val}$ & $F_{auc}$ & $T_{val}$ & $S_{val}$ & $F_{auc}$ \\ \hline
     2x(GAT,GIN) & BN,LRELU,DRPT & 0.808 & 0.884 & 0.947 & \first{0.899} & \first{0.899} & \first{0.929} \\ 
     4xGIN & BN,LRELU,DRPT & 0.657 & 0.810 & 0.898 & 0.849 & 0.850 & 0.914 \\
     4xGAT & BN,LRELU,DRPT & 0.489 & 0.761 & 0.877 & 0.569 & 0.780 & 0.881 \\
     4xGCN & BN,LRELU,DRPT & 0.510 & 0.769 & 0.885 & 0.549 & 0.776 & 0.884 \\ \hline
     2x(GAT,GIN) & BN,LRELU & 0.816 & 0.878 & 0.946 & 0.864 & 0.871 & 0.923 \\
     2x(GAT,GIN) & BN,RELU,DRPT & 0.499 & 0.696 & 0.822 & 0.480 & 0.690 & 0.825 \\
     2x(GAT,GIN) & LRELU,DRPT & \first{0.846} & \first{0.888} & \first{0.950} & 0.839 & 0.866 & 0.919 \\
     2x(GAT,GIN) & RELU & 0.827 & 0.883 & 0.949 & 0.863 & 0.870 & 0.921 \\ \hline
     3xGIN & RELU & 0.633 & 0.821 & 0.910 & 0.661 & 0.820 & 0.894 \\
    \end{tabular}
\end{table}

Experiments covering various encoder designs (Tab.~\ref{tab:ablation_model_encoders}) are divided into three groups. First, we attempt to replace the interleaved layers with a single type repeated four times, preserving the model’s depth. Second, we simplify the normalisation and activation components that follow each neural layer. Finally, we test a substantially simplified encoder with reductions in both aspects. This experiment is aimed to support our unconventional choice of interleaving different types of neural layers.

As shown, removing the interleaved layers generally leads to a drop in performance across all metrics. We hypothesise that the interplay between GAT layers, which prioritise influential neighbourhoods, and GIN layers, which are effective at distinguishing topological structures, reflects the MICM most accurately, thereby enhancing performance. When simplifying the normalisation and activation components, we observe a tendency of the model to overfit to artificial networks drawn from the same distribution as the training data. This naturally leads to poorer generalisation on real-world networks. The smallest drop in performance was observed when dropout alone was removed or when LeakyReLU was replaced with standard ReLU. Finally, we also evaluated a highly simplified encoder consisting of three GIN layers followed by ReLU activations. This configuration showed a substantial performance decline across all metrics, although not the biggest which was observed for encoder build with 4 GCN layers.

\begin{table}[ht!]
    \centering
    \caption{Performance of \textit{ts-net} with different hidden dimensionalities.}
    \label{tab:ablation_model_channels}
    \addtolength{\tabcolsep}{-0.3em}
    \begin{tabular}{l|rrrr|rrrr}
    \multirow{2}{*}{$d_{hidden}$} & \multicolumn{4}{c|}{Artificial networks} & \multicolumn{4}{c}{Real networks} \\
    & $T_{val}$ & $S_{auc}$ & $S_{val}$ & $F_{auc}$ & $T_{val}$ & $S_{auc}$ & $S_{val}$ & $F_{auc}$ \\ \hline
    $200$ & 0.619 & 0.729 & 0.821 & 0.909 & 0.593 & 0.645 & 0.753 & 0.863 \\
    $400$ & \first{0.808} & \first{0.831} & \first{0.884} & \first{0.947} & \first{0.899} & \first{0.814} & \first{0.899} & \first{0.929} \\
    $600$ & 0.796 & 0.806 & 0.861 & 0.931 & 0.886 & 0.794 & 0.877 & 0.925 \\
    $800$ & 0.784 & 0.806 & 0.867 & 0.944 & 0.856 & 0.797 & 0.884 & 0.924 \\
    \end{tabular}
\end{table}

With regards to the next evaluated property of the model, it achieves the best performance with 400 hidden channels. We observe that 200 channels are insufficient, as the model performs worst in this setting. While 600 and 800 channels yield results comparable to 400, they do not outperform it on any metric. Consequently, we adopt 400 as the optimal size.

\begin{table}[ht!]
    \centering
    \caption{Performance of \textit{ts-net} with different aggergation layers.}
    \label{tab:ablation_model_aggregations}
    \addtolength{\tabcolsep}{-0.30em}
    \begin{tabular}{l|rrrr|rrrr}
    \multirow{2}{*}{$aggr(\cdot)$} & \multicolumn{4}{c|}{Artificial networks} & \multicolumn{4}{c}{Real networks} \\
    & $T_{val}$ & $S_{auc}$ & $S_{val}$ & $F_{auc}$ & $T_{val}$ & $S_{auc}$ & $S_{val}$ & $F_{auc}$ \\ \hline
    $min$ & 0.658 & 0.750 & 0.808 & 0.903 & 0.721 & 0.705 & 0.814 & 0.905 \\
    $max$ & 0.678 & 0.803 & 0.873 & 0.942 & 0.743 & 0.750 & 0.851 & 0.915 \\
    $avg$ & 0.721 & 0.787 & 0.859 & 0.940 & 0.799 & 0.737 & 0.841 & 0.919 \\
    $avg_w$ & \first{0.808} & \first{0.831} & \first{0.884} & \first{0.947} & \first{0.899} & \first{0.814} & \first{0.899} & \first{0.929} \\
    $sum$ & 0.655 & 0.723 & 0.799 & 0.897 & 0.679 & 0.692 & 0.806 & 0.888 \\
    $att$ & 0.711 & 0.785 & 0.846 & 0.930 & 0.826 & 0.745 & 0.847 & 0.911 \\
    \end{tabular}
\end{table}

The final architectural experiments (Tab.\ref{tab:ablation_model_aggregations}) focused on the aggregation layer. The highest performance was obtained with the $avg_w$ block proposed by us, followed by the attention-based ($att$) and average ($avg$) aggregators. This outcome supports our intuition: the aggregation mechanism should combine the smoothing properties of averaging with the ability to focus on the most important network's layers.

\subsection{Task}

By conducting this series of experiments, we aimed to evaluate the model’s performance when trained using two loss functions suitable for the considered task. We also sought to examine the expressiveness of the dataset and assess whether training on a data split constructed using spreading outcomes from one regime is sufficient to accurately predict spreading potentials computed under another one.

\begin{table}[ht!]
    \centering
    \caption{Performance of \textit{ts-net} on the considered task under different training loss functions.}
    \label{tab:ablation_task_loss}
    \addtolength{\tabcolsep}{-0.3em}
    \begin{tabular}{l|rrrr|rrrr}
    \multirow{2}{*}{$\mathcal{L}$} & \multicolumn{4}{c|}{Artificial networks} & \multicolumn{4}{c}{Real networks} \\
    & $T_{val}$ & $S_{auc}$ & $S_{val}$ & $F_{auc}$ & $T_{val}$ & $S_{auc}$ & $S_{val}$ & $F_{auc}$ \\ \hline
    ListMLE & \first{0.808} & \first{0.831} & \first{0.884} & \first{0.947} & \first{0.899} & \first{0.814} & \first{0.899} & \first{0.929} \\
    WMAE & 0.758 & 0.801 & 0.871 & 0.938 & 0.887 & 0.789 & 0.867 & 0.919 \\
    \end{tabular}
\end{table}

In the experiment reported in Tab.~\ref{tab:ablation_task_loss}, we compare two loss functions. The first, ListMLE, is designed for ranking prediction tasks. The second, Weighted Mean Absolute Error (WMAE), is primarily used in regression problems and is conceptually close to the task under consideration. As shown, the use of ListMLE consistently yields better performance across all evaluation criteria. Nonetheless, it is worth noting that WMAE does not perform significantly worse.

\begin{table}[ht!]
    \centering
    \caption{Expressiveness of \textit{TopSpreadersDataset} split by $\delta$, as reflected in the generalisation performance of \textit{ts-net}.}
    \label{tab:ablation_task_or-and}
    \addtolength{\tabcolsep}{-0.4em}
    \begin{tabular}{ll|rrrr|rrrr}
    \multirow{2}{*}{Train $\delta$} & \multirow{2}{*}{Test $\delta$} & \multicolumn{4}{c|}{Artificial networks} & \multicolumn{4}{c}{Real networks} \\
    & & $T_{val}$ & $S_{auc}$ & $S_{val}$ & $F_{auc}$ & $T_{val}$ & $S_{auc}$ & $S_{val}$ & $F_{auc}$ \\ \hline
    $AND$& $AND$ & \first{0.808} & \first{0.831} & \first{0.884} & \first{0.947} & \first{0.899} & \first{0.814} & \first{0.899} & \first{0.929} \\
    $OR$ & $AND$ & 0.743 & 0.748 & 0.805 & 0.901 & 0.850 & 0.761 & 0.854 & 0.915 \\ \hline
    $AND$ & $OR$ & \first{0.982} & \first{0.979} & \first{0.993} & \first{0.994} & \first{0.940} & \first{0.912} & 0.946 & 0.954 \\
    $OR$ & $OR$ & 0.955 & 0.957 & 0.976 & 0.982 & 0.924 & 0.911 & \first{0.949} & \first{0.956} \\
    \end{tabular}
\end{table}

We also investigated the dataset’s expressiveness in terms of the model’s ability to generalise. To this end, we trained the \textit{ts-net} on spreading potentials generated using the MICM with one protocol function ($\delta$), and tested it on data from simulations executed with a different protocol. The results are summarised in Tab.~\ref{tab:ablation_task_or-and} and should be interpreted in reference to Fig.~\ref{fig:score_distribution}. As spreading potentials tend to be more evenly distributed when computed from simulations based on the $OR$ protocol, the model trained on such data exhibits inferior generalisation compared to the one trained on data obtained with $\delta = AND$, where the distributions are sharper and super-spreaders are more clearly distinguishable. This supports the claim that identifying super-spreaders using data generated under the $OR$ protocol may be regarded as a trivial task, and therefore unworthy of further investigation.

\subsection{Data transformations}

The final part of the ablation study concerns dataset processing setups. Here, we examine which types of actor features result in the best predictive performance of the model and how transforming the spreading potentials affects this outcome.

\begin{table}[ht!]
    \centering
    \caption{Performance of \textit{ts-net} when trained on different feature vectors.}
    \label{tab:ablation_data_features}
    \addtolength{\tabcolsep}{-0.3em}
    \begin{tabular}{l|rrrr|rrrr}
    \multirow{2}{*}{$x$} & \multicolumn{4}{c|}{Artificial networks} & \multicolumn{4}{c}{Real networks} \\
    & $T_{val}$ & $S_{auc}$ & $S_{val}$ & $F_{auc}$ & $T_{val}$ & $S_{auc}$ & $S_{val}$ & $F_{auc}$ \\ \hline
    $zeros$ & \first{0.808} & \first{0.831} & \first{0.884} & \first{0.947} & \first{0.899} & \first{0.814} & \first{0.899} & \first{0.929} \\
    $ones$ & 0.668 & 0.702 & 0.768 & 0.874 & 0.707 & 0.704 & 0.791 & 0.881 \\
    $central.$ & 0.764 & 0.801 & 0.858 & 0.934 & 0.730 & 0.766 & 0.869 & 0.926 \\
    \end{tabular}
\end{table}

Tab.~\ref{tab:ablation_data_features} presents the impact of different types of actor features. We evaluate three configurations: five-dimensional vectors of ones, five-dimensional zero vectors, and vectors composed of five centrality measures: degree, betweenness, closeness, core number, neighbourhood size — along with VoteRank. All centralities were computed using formulations designed for multilayer networks. Given the nature of the task, we restrict the analysis to these feature sets, since the MICM accounts solely for an actor’s position within the network and no other actor attributes are meaningful in the context of this problem. As shown, the best predictive power is achieved when the model is trained on zero-vectors. This also offers a practical advantage of our proposed approach: to identify super-spreaders, only the network structure is required — incorporating centrality measures does not improve the model’s effectiveness.

The final experiment, albeit one of the most significant, was aimed at identifying the most effective data transformation for the dataset to modify the values of the spreading-potential vector $\mathbf{p}$ (Def.~\ref{def:spv}) in order to facilitate the ranking prediction task. We evaluated seven transformation operators, along with a model trained without any them.

The $norm_{max}$ operator denotes max-normalisation, while $norm_{act}$ corresponds to normalisation by the number of actors in the network --- both applied independently to each coordinate of the vector. The $norm_{act,diam}$ operator applies a combined scheme: $p_{ex}$ and $p_{pi}$ are normalised by the number of actors, whereas $p_{sl}$ and $p_{pl}$ are normalised by the network’s diameter. Next, the $log$ transformation is defined as $\mathbf{p}_{transf} = \log(\mathbf{p} + 1)$. The $scatter$ operator is designed to emphasise the values of actors with high spreading potential while diminishing those of weak spreaders. It is defined as $\mathbf{p}_{transf} = \exp(3 \cdot \mathbf{p}) / \exp(3)$. Finally, we also consider composite transformations, namely $log \circ norm_{max}$ and $scatter \circ norm_{max}$, which apply the corresponding operators in sequence.

\begin{table}[ht]
    \centering
    \caption{Performance of \textit{ts-net} with different data transformations applied.}
    \label{tab:ablation_data_transformations}
    \addtolength{\tabcolsep}{-0.53em}
    \begin{tabular}{l|rrrr|rrrr}
    \multirow{2}{*}{$transf(\cdot)$} & \multicolumn{4}{c|}{Artificial networks} & \multicolumn{4}{c}{Real networks} \\
    & $T_{val}$ & $S_{auc}$ & $S_{val}$ & $F_{auc}$ & $T_{val}$ & $S_{auc}$ & $S_{val}$ & $F_{auc}$ \\ \hline
    $none$ & 0.644 & 0.764 & 0.845 & 0.928 & 0.841 & 0.772 & 0.867 & 0.915 \\
    $norm_{max}$ & 0.808 & \first{0.831} & \first{0.884} & \first{0.947} & 0.899 & 0.814 & \first{0.899} & 0.929 \\
    $norm_{act}$ & 0.648 & 0.725 & 0.807 & 0.902 & 0.811 & 0.714 & 0.810 & 0.905 \\
    $norm_{act,diam}$ & 0.762 & 0.806 & 0.867 & 0.938 & 0.893 & 0.810 & 0.894 & 0.930 \\
    $log$ & 0.631 & 0.735 & 0.816 & 0.908 & 0.800 & 0.698 & 0.800 & 0.899 \\
    $log \circ norm_{max}$ & 0.648 & 0.711 & 0.786 & 0.892 & 0.731 & 0.712 & 0.801 & 0.892 \\
    $scatter$ & \first{0.850} & 0.826 & 0.881 & \first{0.947} & \first{0.921} & \first{0.826} & 0.897 & \first{0.934} \\
    $norm_{max}\circ scatter$ & 0.633 & 0.749 & 0.832 & 0.919 & 0.644 & 0.744 & 0.857 & 0.905 \\
    \end{tabular}
\end{table}

As shown in Tab.~\ref{tab:ablation_data_transformations}, two transformations yielded the best results: $norm_{max}$ and $scatter$. However, due to its superior generalisation ability on real-world networks and the possibility of inverting the transformation, we select the latter for the final comparison with competitive methods. Among the remaining operators, $norm_{act,diam}$ also achieved solid performance, with the additional advantage of enabling the recovery of non-relative values for the predicted spreading-potential vector.

\section{Evaluation on other metrics}
\label{app:eval_other_metrics}

In addition to evaluating the proposed model against competitive baselines using the relative cumulative score (Def.~\ref{def:sps-relcum}), we also employed metrics commonly used in the domain of ranking-learning. Instead of assessing prediction quality based on the scores assigned to top-$k$ ranked actors, these metrics evaluate only the alignment between the indices in $\mathbf{\hat{R}}$ and $\mathbf{R}$. Specifically, we report:

\begin{itemize}
  \item Precision — the fraction of correctly predicted actors in $\mathbf{\hat{R}}$;
  \item Jaccard index — the overlap between the sets $\mathbf{\hat{R}}$ and $\mathbf{R}$.
\end{itemize}

We report these metrics, similarly to the main metrics used in the study (Tab.~\ref{tab:metrics}), at three levels of the ranking: at its top position ($T$), at the cut-off separating super-spreaders ($S$), as illustrated in Fig.~\ref{fig:score_distribution}, and over the full ranking ($F$). For each metric, we report the average score computed across all prefixes of the ranking, that is, over all top-$k$ sets for $k = 1, 2, \dots, \{T, S, F\}$.

\begin{table}[ht!]
    \centering
    \caption{Average \textbf{precision} of the evaluated methods across two network splits. Predictions with the highest quality ($\sim1$) are highlighted in green, with darker shades indicating better ranks.}
    \label{tab:comparison_ap}
    \begin{tabular}{l|rrr|rrr}
    \multirow{2}{*}{$\phi$} & \multicolumn{3}{c|}{Artificial networks} & \multicolumn{3}{c}{Real networks} \\
     & $T$ & $S$ & $F$ & $T$ & $S$ & $F$ \\ \hline
    \textit{random} & 0.004 & 0.087 & 0.502 & 0.010 & 0.092 & 0.505 \\ \hline
    \textit{deg-c} & \first{0.250} & \second{0.370} & \third{0.629} & \first{0.143} & \third{0.426} & \third{0.612} \\
    \textit{deg-cd} & \first{0.250} & \second{0.370} & \second{0.631} & \first{0.143} & \second{0.427} & \second{0.621} \\
    \textit{nghb-s} & 0.062 & 0.250 & 0.582 & \first{0.143} & 0.394 & 0.579 \\
    \textit{nghb-sd} & 0.000 & 0.249 & 0.577 & \first{0.143} & 0.382 & 0.587 \\ \hline
    \textit{deep-im} & 0.000 & 0.098 & 0.506 & 0.000 & 0.095 & 0.497 \\
    \textit{mn2v-km} & 0.000 & 0.147 & 0.547 & 0.000 & 0.138 & 0.533 \\
    \textit{\textbf{ts-net}} & \first{0.250} & \first{0.382} & \first{0.638} & \first{0.143} & \first{0.450} & \first{0.624} \\
    \end{tabular}
\end{table}

\begin{table}[ht!]
    \centering
    \caption{Average \textbf{Jaccard} index of the evaluated methods across two network splits. Predictions with the highest quality ($\sim1$) are highlighted in green, with darker shades indicating better ranks.}
    \label{tab:comparison_jaccard}
    \begin{tabular}{l|rrr|rrr}
    \multirow{2}{*}{$\phi$} & \multicolumn{3}{c|}{Artificial networks} & \multicolumn{3}{c}{Real networks} \\
    & $T$ & $S$ & $F$ & $T$ & $S$ & $F$ \\ \hline
    \textit{random} & 0.004 & 0.047 & 0.389 & 0.010 & 0.053 & 0.392 \\ \hline
    \textit{deg-c }& \first{0.250} & \second{0.242} & \third{0.493} & \first{0.143} & \second{0.289} & \third{0.477} \\
    \textit{deg-cd} & \first{0.250} & \third{0.241} & \second{0.494} & \first{0.143} & \second{0.289} & \first{0.488} \\
    \textit{nghb-s} & 0.062 & 0.150 & 0.451 & \first{0.143} & 0.263 & 0.445 \\
    \textit{nghb-sd} & 0.000 & 0.149 & 0.446 & \first{0.143} & 0.251 & 0.454 \\ \hline
    \textit{deep-im} & 0.000 & 0.053 & 0.391 & 0.000 & 0.052 & 0.385 \\
    \textit{mn2v-km} & 0.000 & 0.084 & 0.424 & 0.000 & 0.080 & 0.415 \\
    \textit{\textbf{ts-net}} & \first{0.250} & \first{0.254} & \first{0.503} & \first{0.143} & \first{0.308} & \first{0.488} \\
    \end{tabular}
\end{table}


The results, averaged across the tested networks, are presented in Tab.~\ref{tab:comparison_ap} and Tab.~\ref{tab:comparison_jaccard}. 
It is worth noting that the values for the $T$ scenario are consistent across all metrics, as expected. 

Notably, despite the different evaluation criteria employed, our method consistently achieves the top rank across all metrics. 
This performance aligns with the conclusions drawn in the main part of the paper and further underscores the robustness of the proposed model. At the same time, this evaluation methodology highlights the inherent difficulty of identifying super-spreaders in multilayer networks. The values obtained by all evaluated methods underscore both the complexity of the problem and the substantial room for further methodological advancements.

\section{Scalability of the model}
\label{app:time_complexity}

To assess the scalability of the proposed model in terms of computational efficiency, two aspects were analysed: GPU memory consumption and processing time as a function of the network size. The evaluation was based on the test set, as it was also processed by the competing methods considered in this study.

\begin{figure}[ht!]
    \centering
    \includegraphics[width=.8\linewidth]{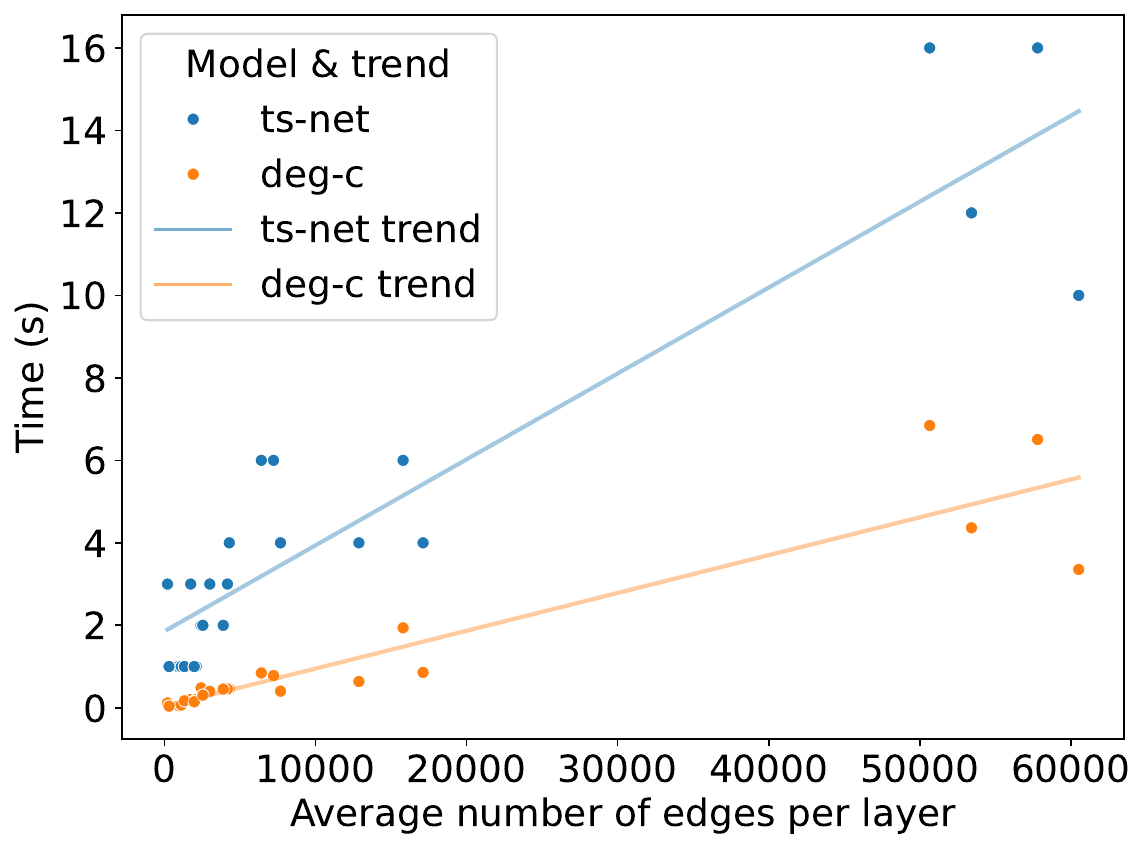}
    \caption{Inference time of \textit{ts-net} compared to that of \textit{deg-c}.}
    \label{fig:time_comparison}
\end{figure}

First, Fig.~\ref{fig:time_comparison} presents processing times depending on the average number of edges per network layer for \textit{ts-net} and, for reference, the light yet effective \textit{deg-c}. The results for the proposed model indicate that processing time increases almost linearly, from near-zero values for small networks to approximately $16$ seconds for networks containing over $60,000$ edges per layer. In the case of \textit{deg-c}, the scaling is also linear but with a gentler slope. Linear regression trends confirm these observations: \textit{ts-net} exhibits a slope of $0.0002$ with a strong fit ($R^2 = 0.853$), while \textit{deg-c} shows a slightly lower slope of $0.00009$ with $R^2 = 0.860$. It is worth noting that comparisons with \textit{deg-c} on larger graphs would not be representative due to presumably lower-level optimisation mechanisms in the \texttt{degree} function from the NetworkX library, which underpins this heuristic.

Regarding the second aspect, thanks to the use of \texttt{NeighborLoader}, the GPU memory usage of \textit{ts-net} during inference remains constant at around $15.16$ GB across the entire test set. This enables effective scaling and ensures applicability to real-world scenarios involving continuously growing networks.

\section{Training details}

All \textit{ts-net} variants were trained using the \texttt{AdamW} optimiser with a \texttt{OneCycleLR} learning rate scheduler. The implementation was developed in Python 3.12 using PyTorch 2.3.1. Training was limited to a maximum of $50$ epochs, with early stopping applied and a patience threshold of $30$ epochs. \texttt{NeighborLoader} with a batch size of $128$, using an induced $N$-hop neighbourhood sampling strategy aligned with the number of convolutional layers was used. Specifically, up to $32$ neighbours were sampled at the first hop, with the number halved at each subsequent hop. The maximum learning rate was set to $0.003$. All experiments involving \textit{ts-net} were conducted on a machine equipped with an Intel(R) Xeon(R) Gold 6330 CPU @ 2.00GHz, 64~GB RAM, and an NVIDIA Titan RTX GPU. Hyperparameters for baseline machine learning methods were configured as described in their respective original papers and trained on CPU due to high memory requirements.

\end{document}